\documentclass[%
 reprint,
 amsmath,amssymb,
 aps,
floatfix,
]{revtex4-2}
\usepackage{placeins}
\usepackage{graphicx}
\usepackage{dcolumn}
\usepackage{bm}
\usepackage{gensymb}
\usepackage{makecell}
\usepackage[caption=false]{subfig}
\usepackage{hyperref}
\usepackage[T1]{fontenc}
\usepackage[utf8]{inputenc}
\newcommand{\subfigimg}[3][,]{%
  \setbox1=\hbox{\includegraphics[#1]{#3}}
  \leavevmode\rlap{\usebox1}
  \rlap{\hspace*{0pt}\raisebox{\dimexpr\ht1-2\baselineskip}{#2}}
  \phantom{\usebox1}
}

\begin{document}

\preprint{}

\title{Impact of the 50~Hz harmonics on the beam evolution of the Large Hadron Collider}

\thanks{Research supported by the HL-LHC project.}

\author{S.~Kostoglou}
\email{sofia.kostoglou@cern.ch} 
\affiliation{%
 CERN, Geneva 1211, Switzerland \\
}%
\affiliation{%
 National Technical University of Athens, Athens 15780, Greece\\
}%

\author{G.~Arduini}%
\affiliation{%
 CERN, Geneva 1211, Switzerland \\
}%

\author{L.~Intelisano}%
\affiliation{%
 CERN, Geneva 1211, Switzerland \\
}%
\affiliation{%
 INFN, Sapienza  Universit\`a di Roma, Rome 00185, Italy\\
}%

\author{Y.~Papaphilippou}%
\affiliation{%
 CERN, Geneva 1211, Switzerland \\
}%
 \author{G.~Sterbini}%
\affiliation{%
 CERN, Geneva 1211, Switzerland \\
}%

\date{\today}

\begin{abstract}

Harmonics of the mains frequency (50~Hz) have been systematically observed in the transverse beam spectrum of the Large Hadron Collider (LHC) since the start of its operation in the form of dipolar excitations. In the presence of strong non-linearities such as beam-beam interactions, as many of these power supply ripple tones reside in the vicinity of the betatron tune they can increase the tune diffusion of the particles in the distribution, leading to proton losses and eventually to a significant reduction of the beam lifetime. The aim of this paper is to determine whether the 50~Hz harmonics have an impact on the beam performance of the LHC. A quantitative characterization of the ripple spectrum present in the operation of the accelerator, together with an understanding of its source is an essential ingredient to also evaluate the impact of the 50~Hz harmonics on the future upgrade of the LHC, the High Luminosity LHC (HL-LHC). To this end, simulations with the single-particle tracking code, SixTrack, are employed including a realistic ripple spectrum as extracted from experimental observations to quantify the impact of such effects in terms of tune diffusion, Dynamic Aperture and beam lifetime. The methods and results of the tracking studies are reported and discussed in this paper.

\end{abstract}

\maketitle

\section{Introduction}
Power supply ripple at harmonics of the mains power frequency (50~Hz) has been systematically observed in the transverse beam spectrum of the Large Hadron Collider (LHC) since the start of its operation. In a previous paper \cite{previous}, we investigated the source of the perturbation based on a systematic analysis of experimental observations. It was clearly shown that the 50~Hz harmonics are not an artifact of the instrumentation system but correspond to actual beam oscillations. 

Two regimes of interest have been identified in the transverse beam spectrum: a series of 50~Hz harmonics extending up to approximately 3.6~kHz, referred to as the \textit{low-frequency cluster}, and a cluster of 50~Hz harmonics centered around the alias of the betatron frequency, i.e., $f_{\rm rev}-f_x$, where $f_{\rm rev}$=11.245~kHz is the LHC revolution frequency and $f_x$ the betatron frequency, namely the \textit{high-frequency cluster}. Both clusters are the result of dipolar beam excitations and it was demonstrated with dedicated experiments that the source of the low-frequency cluster are the eight Silicon Controlled Rectifier (SCR) power supplies of the main dipoles \cite{SCR, SCR2}. Table~\ref{tab:50Hz} presents the frequency and the amplitude of the 12 most important 50~Hz harmonics in the transverse spectrum of Beam 1 during collisions as observed in the Q7 pickup of the transverse damper Observation Box (ADTObsBox) \cite{ADT, ADT3, ADT4}. The offsets induced in Beam 2 due to the power supply ripple are also depicted, which are approximately lower by a factor of two compared to Beam 1. Based on the source, the 50~Hz harmonics are expected to be present in the future operation of the accelerator.

\begin{table*}
\caption{\label{tab:50Hz}%
The beam offsets observed in the Q7 pickup of the transverse damper for the 12 largest 50~Hz harmonics in the horizontal spectrum of Beam 1 (B1) and 2 (B2) during collisions.}
\begin{ruledtabular}
\begin{tabular}{cccccccc}
\textrm{Frequency (kHz)}& Cluster &
\textrm{B1 offset (nm)}  & \textrm{B2 offset (nm)} & \textrm{Frequency (kHz)}& Cluster &
\textrm{B1 offset (nm)}  & \textrm{B2 offset (nm)}\\
\colrule
7.6 & High& 166.2 & 76.51 & 7.7 & High& 109.75 & 51.38\\
1.2 & Low& 98.63 & 71.87 &  7.65 & High& 97.4 & 51.97 \\
7.8 & High& 92.75 & 42.78 & 2.95 & Low& 91.62 & 32.21\\
2.35 & Low& 88.48 & 29.61 &  7.9 & High& 82.23 & 25.46 \\
7.5 & High& 77.82 & 35.08 & 2.75 & Low& 67.43 & 31.77 \\
7.85& High & 67.37 & 18.86 & 0.6 & Low &54.14 & 14.33  \\

\end{tabular}
\end{ruledtabular}
\end{table*}

In the presence of strong non-linearities such as beam-beam interactions and fields of non-linear magnets, a modulation in the dipole strength due to power supply ripple results in a resonance condition \cite{tomas1, tomas2}:
\begin{equation}
\label{eq:tune}
    l \cdot Q_x +  m\cdot Q_y +  p \cdot Q_p = n
\end{equation}
where $l, m,p,n$ are integers, $Q_x$ and $Q_y$ are the horizontal and vertical tunes, respectively, and $Q_p$ is the power supply ripple tune. Power supply ripple can impact the beam performance by introducing resonances in addition to the ones excited due to the non-linearities of the lattice. These resonances can increase the tune diffusion of the particles in the distribution, which may lead to the reduction of the beam lifetime. The aim of the present paper is to determine whether the 50~Hz harmonics act as a limiting factor to the beam performance of the LHC, as well as its future upgrade, the High-Luminosity LHC (HL-LHC) \cite{Apollinari:2284929}, using single-particle tracking simulations with power supply ripple in the form of dipolar excitations. The impact of power supply ripple that results in a tune modulation is discussed in a previous paper \cite{kostoglou2020tune}. 

First, the motion of a particle in the presence of a modulated dipolar field error is described using a linear formalism (Section~\ref{app:modulated_dipole}). Second, single-tone dipolar excitations are considered in the tracking simulations to define a minimum power supply ripple threshold that results in a reduction of the Dynamic Aperture (DA) in the presence of strong non-linearities such as long-range and head-on beam-beam encounters, chromatic arc sextupoles and Landau octupoles. Then, the impact of the harmonics on the beam performance in terms of tune diffusion, DA and lifetime is discussed in Sec.~\ref{Sec:simulations} by including in the simulations a power supply ripple spectrum as acquired from experimental observations. Finally, the impact of controlled dipolar excitations on the beam lifetime, conducted with the transverse damper, is described in Section~\ref{Sec:adt_excitations}, which provides a tool for the validation of the DA simulations in the presence of power supply ripple.

\section{Linear formalism for a modulated dipolar field error}
\label{app:modulated_dipole}

In a circular accelerator, the kick related to a modulated dipolar field error \(\Theta_p\) with a deflection \(\theta_{p}\) and a frequency of \(Q_{p}\) oscillations per turn can be represented in normalized phase space as:
\begin{equation}
    \bar{P}_n = \begin{pmatrix} 0 \\ \Theta_{p}(n) \end{pmatrix} = \begin{pmatrix} 0 \\ \sqrt{\beta} \theta_{p} \cos{(2 \pi Q_p n)} \end{pmatrix},
\label{appendix:eq_1}
\end{equation}
where \(n\) is the turn considered and $\beta$ is the $\beta$-function at the location of the perturbation.

In the linear approximation, considering only the horizontal motion and assuming that the source and the observation point are situated in the same location, the vector representation of the position and momentum at turn \(N\) is:
\begin{equation}
\bar{X}_N= 
\left(\begin{array}{c}
\bar{x}_N
\\
\bar{x}_N'
\end{array}\right)=
\sum_{n=-\infty}^{N}M^{N-n}\bar{P}_n,
\label{appendix:eq_2}
\end{equation}
where \(M\) is the linear rotation equal to:
\begin{equation}
    M^{N}=\left(\begin{array}{cc}{\cos (2 \pi Q N)} & {\sin (2 \pi Q N)} \\ {-\sin (2 \pi Q N)} & {\cos (2 \pi Q N)}\end{array}\right)
\label{appendix:eq_3}
\end{equation}
with \(Q\) representing the machine betatron tune. Combining Eq.~\eqref{appendix:eq_1}, \eqref{appendix:eq_2} and \eqref{appendix:eq_3} and assuming that the perturbation is present from $n \rightarrow-\infty$, yields:
\begin{align}
 & \bar{x}_N=  \sqrt{\beta} \theta_{p} \sum_{n=-\infty}^{N} \cos{(2 \pi Q_p n)} \sin{\left(2 (N-n) \pi Q\right)},  \nonumber \\
     &  \bar{x}'_N=  \sqrt{\beta} \theta_{p}  \sum_{n=-\infty}^{N}  \cos{(2 \pi Q_p n)} \cos{\left(2 (N-n) \pi Q \right)}.
\label{appendix:eq_4}
\end{align}
In Eq.~\eqref{appendix:eq_4}, using Euler's formula and writing the sum expressions in terms of geometric series yields:

\begin{widetext}
\begin{equation}
{
 \begin{aligned}
{\bar x}_N  =\sqrt{\beta} \theta_p \lim_{k\to-\infty} \Bigr(&\frac{\cos (2 \pi Q_p N) \sin(2 \pi Q)}{2(\cos(2\pi Q_p)-\cos(2\pi Q))}  \\
            &+\frac{- \cos (2 \pi Q_p (k-1) ) \sin (2 \pi Q (k - N)) +  \cos (2 \pi  Q_p k ) \sin (2 \pi Q(k - 1 - N) )}{2(\cos(2\pi Q_p)-\cos(2\pi Q))} \Bigl) \;\;,\\
{\bar x'}_N = \sqrt{\beta} \theta_p \lim_{k\to-\infty} \Bigr(&\frac{\cos (2 \pi Q_p (N+1) )-\cos (2 \pi Q_p N )\cos (2 \pi  Q)}{2(\cos{(2\pi Q_p)}-\cos{(2\pi Q)})}\\
&+\frac{\cos (2 \pi  Q_p (k-1)) \cos (2 \pi  Q (k-N))+\cos (2 \pi Q_p k)  \cos (2 \pi  Q (k-1-N))}{2(\cos{(2\pi Q_p)}-\cos{(2\pi Q)})} \;\;.\\
\label{eq:sum2}
\end{aligned}}
\end{equation}
\end{widetext}
The limit of the terms in the numerator of Eq.~\eqref{eq:sum2} involving $k$ in the argument of the trigonometric functions is bounded but indeterminate. Using  the Ces\`aro mean the trigonometric products involving $k$ as an argument in Eq.~\eqref{eq:sum2} is equal to zero and the expressions for the position and momentum simply become:
\begin{align}
\label{eq:sumsol}
{\bar x}_N  &= \sqrt{\beta} \theta_p \frac{\cos (2 \pi Q_p N) \sin(2 \pi Q) }{2(\cos{(2\pi Q_p)}-\cos{(2\pi Q)})}, \\
{\bar x'}_N &= \sqrt{\beta} \theta_p \frac{\cos(2 \pi Q_p N) \cos(2 \pi Q)-\cos (2 \pi  (N+1) Q_p)}{2(\cos{(2\pi Q_p)}-\cos{(2\pi Q)})}. \nonumber
\end{align}
In physical coordinates, the maximum offset computed from Eq.~\eqref{eq:sumsol} is:

\begin{equation}
    |x_{\text{max}, N}|=\left\lvert \frac{\beta \theta_{p}\sin{(2\pi  Q)}}{2 \left( \cos{(2\pi Q_{p})}-\cos(2\pi Q)  \right) }\right\rvert.
     \label{appendix:eq_6}
\end{equation}

\section{Simulations of 50~Hz harmonics with beam-beam interactions}
\label{Sec:simulations}

Tracking simulations are performed using the single-particle symplectic code SixTrack \cite{sixtrack, sixtrack2}. A distribution of particles is tracked in the element-by-element LHC and HL-LHC lattice at top energy, including important non-linearities such as head-on and long-range beam-beam encounters, sextupoles for chromaticity control and Landau octupoles for the mitigation of collective instabilities. For the LHC case, the chromaticity in the horizontal and vertical plane is $Q'$=15 and the amplitude detuning coefficients $\alpha_{n_1 n_2}=\partial Q_{n_1} / \partial(2 J_{n_2})$ with $n_1, n_2 \in \{x,y\}$ are:
\begin{eqnarray}
    \alpha_{\rm xx}=16.8\times 10^{4} \ \text{m}^{-1} , \alpha_{\rm xy} = -11.8\times 10^{4} \ \text{m}^{-1},  \nonumber \\
    \alpha_{\rm yy} =16.5\times 10^{4} \ \text{m}^{-1} 
\end{eqnarray}
as computed from the PTC-normal module of the MAD-X code \cite{madx}.

The beam-beam interactions are simulated with the weak-strong approximation. The maximum duration of the tracking is \(\rm 10^6\) turns that corresponds to approximately 90 seconds of beam collisions in operation. The main parameters used in simulations for LHC and HL-LHC are presented in Table~\ref{tab:table_parameters_collision}. For the LHC, the simulated conditions refer to the start of the collisions. The simulated conditions in the HL-LHC case correspond to the nominal operational scenario at the end of the $\beta^*$-leveling, taking place after several hours from the start of the collisions \cite{HL_LHC_scenarios}.

\begin{table*}
\caption{\label{tab:table_parameters_collision}%
The LHC and HL-LHC parameters at top energy used in the tracking simulations. The LHC parameters correspond to the start of collisions, while the HL-LHC parameters refer to the end of the $\beta^*$-leveling as envisaged in the nominal scenario \cite{HL_LHC_scenarios}. 
}
\begin{ruledtabular}
\begin{tabular}{lcc}
\textrm{Parameters (unit)}&
\textrm{LHC (values)} & \textrm{HL-LHC (values)}\\
\colrule
Beam energy (TeV) & 6.5 & 7\\
Bunch spacing (ns) & 25 & 25\\
RMS bunch length (cm) & 7.5 & 7.5 \\
Bunch population (protons)  & $1.25 \times 10^{11}$ & $1.2\times 10^{11}$\\
Beam-beam parameter $\xi_{x,y} $ & 7.6$\times 10^{-3}$ & 5.8$\times 10^{-3}$ \\
Horizontal tune \(Q _x\) &0.31 & 0.31/0.315\footnote{\label{note1}Nominal/optimized working point.} \\ 
Vertical tune \(Q_y\) &0.32/0.315\textsuperscript{\ref{note1}}  & 0.32\\
Horizontal and vertical normalized emittance (\(\rm \mu m \ rad\)) & 2.0 & 2.5\\
Horizontal and vertical chromaticity  & 15 & 15\\
Octupole current (A)  & 550 & -300\\
IP1/5 Half crossing angle (\(\rm \mu rad\)) & 160 &  250\\
Horizontal and vertical IP1/5 \(\beta^*\) (cm) & 30 & 15\\ 
Total RF voltage (MV) & 12 & 16 \\
Synchrotron frequency (Hz) & 23.8 & 23.8 \\
Revolution frequency (kHz) & 11.245 & 11.245\\
Relative momentum deviation $\delta p /p$ & 27$\times 10^{-5}$ & 27$\times 10^{-5}$ \\
Horizontal \(\beta\)-function at Q7 pickup (m) & 105.4 & 102.3\\ 
\end{tabular}
\end{ruledtabular}
\end{table*}

In a previous paper \cite{previous}, it was demonstrated that, as far as the low-frequency cluster is concerned, the power supply ripple is distributed in all eight sectors. An accurate representation of the power supply ripple propagation across the chains of the LHC dipoles requires a model of the transfer function as a transmission line for all the spectral components in the low-frequency cluster, similarly to the studies performed for the SPS \cite{Burla:263879}. Furthermore, as the exact mechanism of the high-frequency cluster is not yet clearly identified, an accurate transfer function is not known at the moment.

To simplify these studies, the distributed power supply ripple is projected to a single location with an equivalent impact on the beam's motion. Specifically, a horizontal modulated dipolar source is included at the location of the Q7 pickup of the transverse damper, which coincides with the observation point. To simulate the dipolar excitation, the strength of a horizontal kicker is modulated with a sinusoidal function:
\begin{equation}
\label{eq:mod}
    \Delta k (t) = \theta_p \cos{(2\pi f_p t)},
\end{equation}
where $t$ is the time, $\theta_p$ the deflection,  $f_p=Q_p \cdot f_{\rm rev}$ the power supply ripple frequency and $f_{rev}$ the revolution frequency listed in Table~\ref{tab:table_parameters_collision}. In this way, the contribution of all the dipoles is represented by a single, equivalent kick and the offsets observed in the LHC spectrum are reproduced in the simulations. The maximum offset induced on the particle's motion is computed from Eq.~\eqref{appendix:eq_6} with the horizontal $\beta$-function listed in Table~\ref{tab:table_parameters_collision}.

The simulations are then repeated for the HL-LHC case. The need to perform projections for the HL-LHC is justified by the fact that no modifications are envisaged for the power supplies of the main dipoles. Consequently, based on the source, the 50~Hz harmonic are expected to also be present in the HL-LHC era. In the following sections, the HL-LHC studies are based on the power supply ripple spectrum acquired experimentally from LHC, although it is expected that the foreseen upgrade of the transverse damper system will lead to a more efficient suppression of the harmonics.

\subsection{Impact of single-tone ripple spectrum on the Dynamic Aperture}

As a first step, we consider individual tones in the power supply ripple spectrum for increasing values of the deflection $\theta_p$. For each study, a different combination of the frequency $f_p$ and the amplitude of the excitation $\theta_p$ is selected. For each case, the minimum DA in the initial configuration space, i.e., the initial horizontal $x$ and vertical $y$ displacements is compared to the value derived in the absence of power supply ripple. In these simulations, the tune modulation due to the coupling of the synchrotron oscillations to the betatron motion for a chromaticity and a relative momentum deviation listed in Table~\ref{tab:table_parameters_collision} is also included. 

The scan in the ripple parameter space (\(f_p, \theta_p\)) is performed to define the most dangerous tones of the low and high-frequency cluster, i.e., the frequencies that, for a constant excitation amplitude, have a maximum impact on the DA. Then, the offset induced in the beam motion at the location of observation point is estimated from Eq.~\eqref{appendix:eq_6} with the parameters $f_p$, $\theta_p$ and the $\beta$-function of Table~\ref{tab:table_parameters_collision}. A threshold for the minimum offset at the Q7 pickup that leads to a reduction of DA is determined. 

\begin{figure}
\includegraphics[width = \columnwidth]{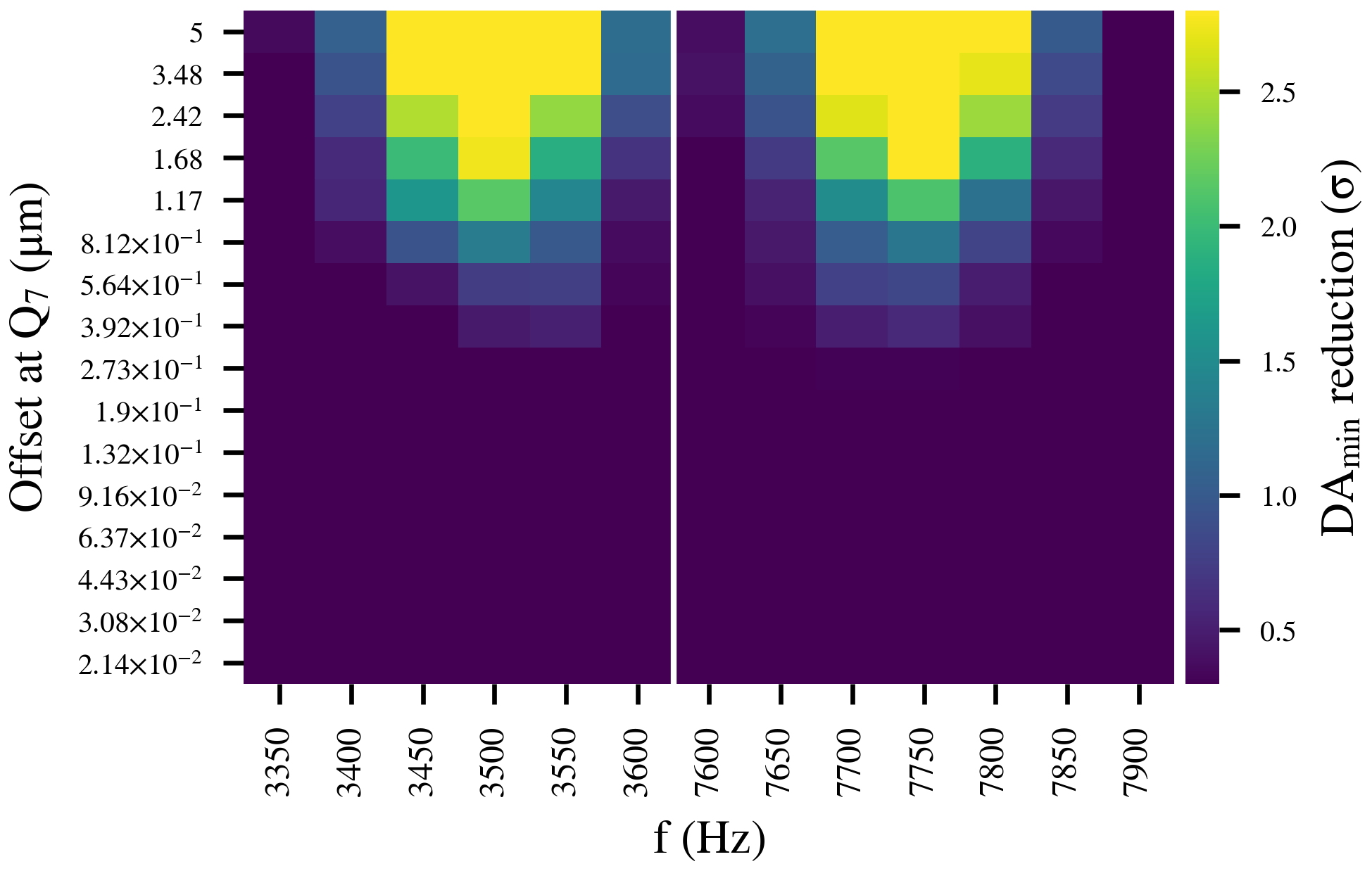}
\caption{\label{fig:heatmap} The ripple frequency as a function of the horizontal offset in the Q7 pickup, as computed from the deflection and Eq.~\eqref{appendix:eq_6}. A color code is assigned to the reduction of the minimum DA from the value computed in the absence of power supply ripple.}
\end{figure}

Figure~\ref{fig:heatmap} presents the ripple frequency as a function of the offset at the observation point. The harmonics of the low and the high-frequency cluster that reside in the vicinity of \(f_x\) and \(f_{\rm rev}-f_x\) have been selected for the analysis. A color code is assigned to the reduction of the minimum DA to illustrate the regimes with a negligible (blue) and significant (yellow) impact.

 From the scan, it is evident that the ripple frequencies with the largest impact are the ones that reside in the proximity of the tune and its alias. For the LHC, an offset threshold of \(\rm 0.4 \ \mu m\) is defined, while this limit reduces to \(\rm 0.2 \ \mu m\) for the HL-LHC. For comparison, the maximum excitation observed experimentally due to the 50~Hz lines is approximately 0.16~\(\rm \mu m\) as presented in Table~\ref{tab:50Hz}.



\subsection{Frequency Map Analysis with a realistic power supply ripple spectrum}

A more significant impact on the beam performance is anticipated in the presence of multiple ripple tones, similar to the experimental observations of the 50~Hz harmonics, than considering individual tones due to the resonance overlap \cite{intro1994_1, SPS2_tune}. To this end, a realistic power supply ripple spectrum must be included in the simulations. Similarly to Table~\ref{tab:50Hz}, the offsets and the frequencies of the 40 largest 50~Hz harmonics are extracted from the horizontal beam spectrum as observed during operation. For each frequency, the equivalent kick at the location of the Q7 pickup is computed from Eq.~\eqref{appendix:eq_6} and it used as an input for the power supply ripple simulations.

\begin{figure}
\includegraphics[width = \columnwidth]{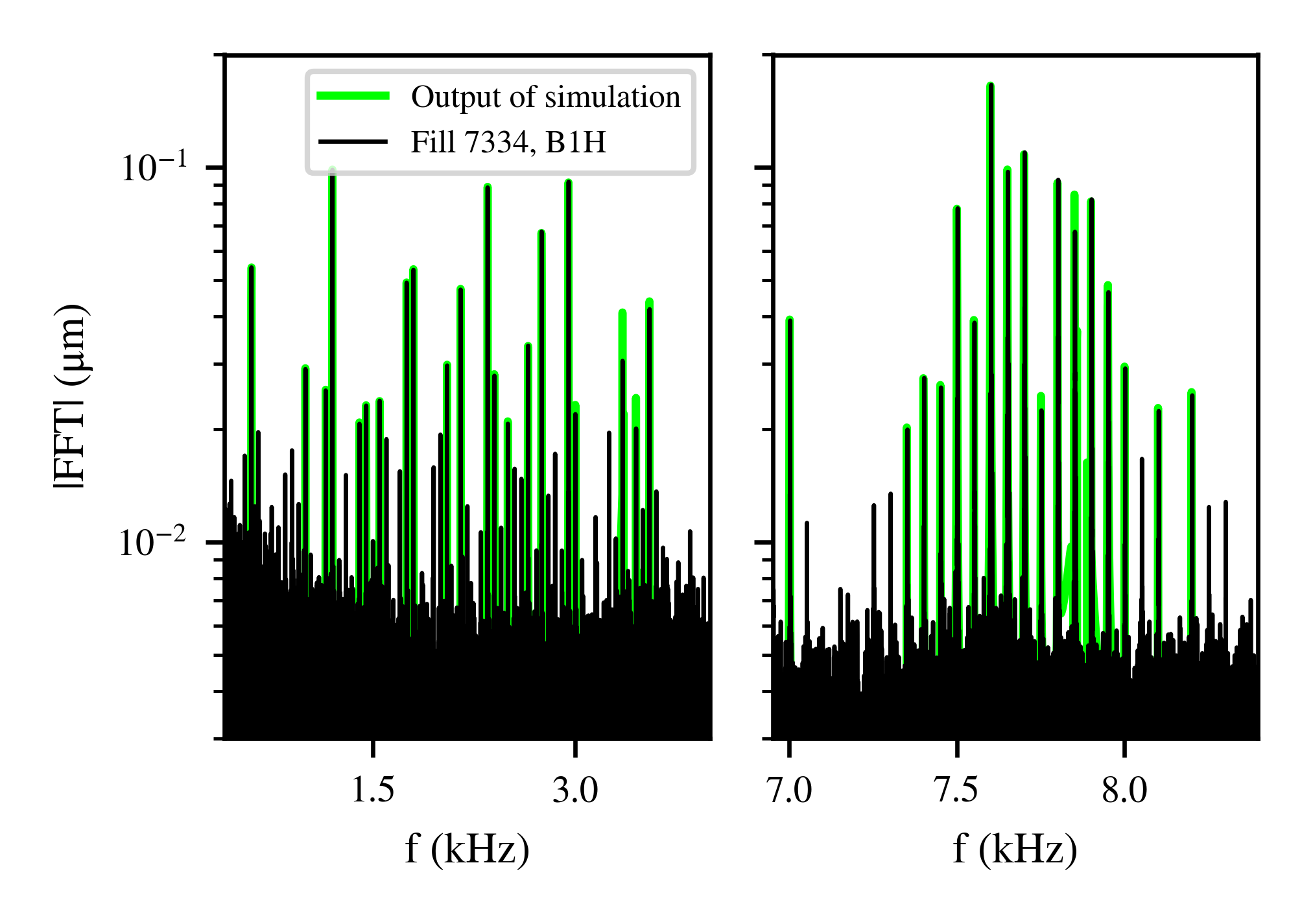}
\caption{\label{fig:spectrum} The spectrum centered around the low (left) and high (right) frequency cluster as acquired experimentally from the horizontal plane of Beam 1 (black) and from tracking simulations (green).}
\end{figure}

Figure~\ref{fig:spectrum} shows the horizontal spectrum of Beam 1 from the experimental observations (black) centered around the low (left) and high (right) frequency cluster. A single particle is tracked in the LHC lattice including the most important 50~Hz harmonics. The output of the simulations (green) is also illustrated in Fig.~\ref{fig:spectrum}. The comparison of the two is a sanity check illustrating the good agreement between the simulated and the experimental beam spectrum. A similar agreement (between simulated and expected beam spectrum) is found for the HL-LHC case, were the equivalent kicks have been recomputed due to a small variation of the \(\beta\)-function as shown in Table~\ref{tab:table_parameters_collision}.

The studies are organized as follows. First, a study in the absence of power supply ripple is performed that is used as a baseline. Second, the largest 50~Hz harmonics of the low-frequency cluster are considered. Then, a separate study is conducted including only the most important harmonics of the high-frequency cluster. Last, both regimes are included in the simulations.

 Particles extending up to 6 \(\rm \sigma\) in the initial configuration space are tracked for $\rm 10^4$ turns in the LHC and HL-LHC lattice. The particles are placed off-momentum as shown in Table~\ref{tab:table_parameters_collision}, without however considering the impact of synchrotron oscillations. A Frequency Map Analysis (FMA) is performed for each study \cite{fma1, fma2, fma3, fma4}. The turn-by-turn data are divided into two groups containing the first and last 2000 turns, respectively. The tune of each particle is computed for each time interval using the Numerical Analysis of Fundamental Frequencies (NAFF) algorithm \cite{NAFF1, NAFF, PhysRevAccelBeams.22.071002}. Comparing the variation of the tune of each particle across the two time spans reveals information concerning its tune diffusion rate $D$ that is defined as:
 
\begin{equation}
\label{eq:diffusion}
    D = \sqrt{ (Q_{x_1}- Q_{x_2})^2 + (Q_{y_1}- Q_{y_2})^2 },
\end{equation}
 where $Q_{j_1}, Q_{j_2}$ with $j \in \{x,y\}$ denote the tunes of each particle in the first and second time interval, respectively. 

\begin{figure*}
\subfloat{\subfigimg[width = 0.49\textwidth]{ \textbf{a)}}{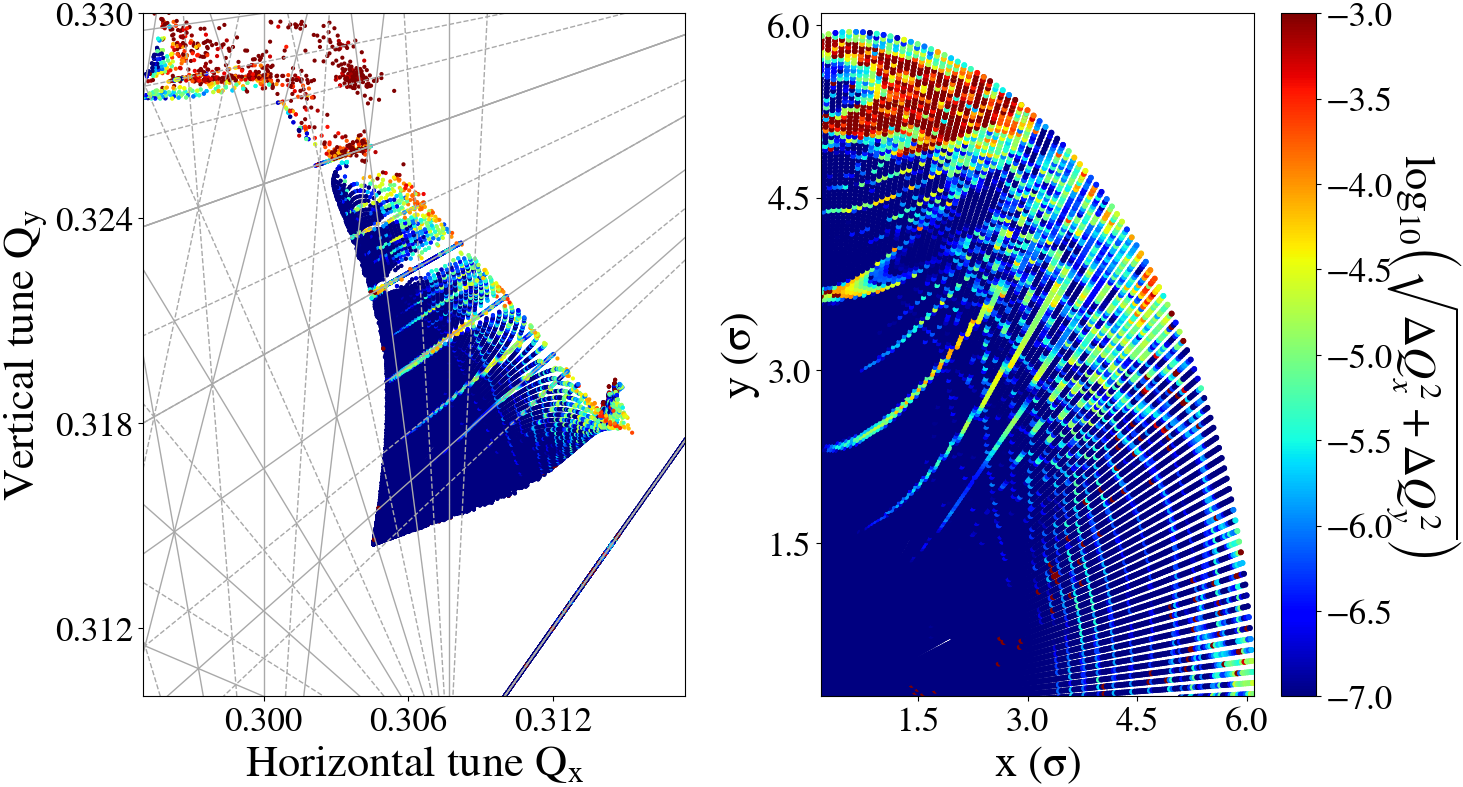} \label{subfig:FMA_without}} 
\subfloat{\subfigimg[width = 0.49\textwidth]{\textbf{b)}}{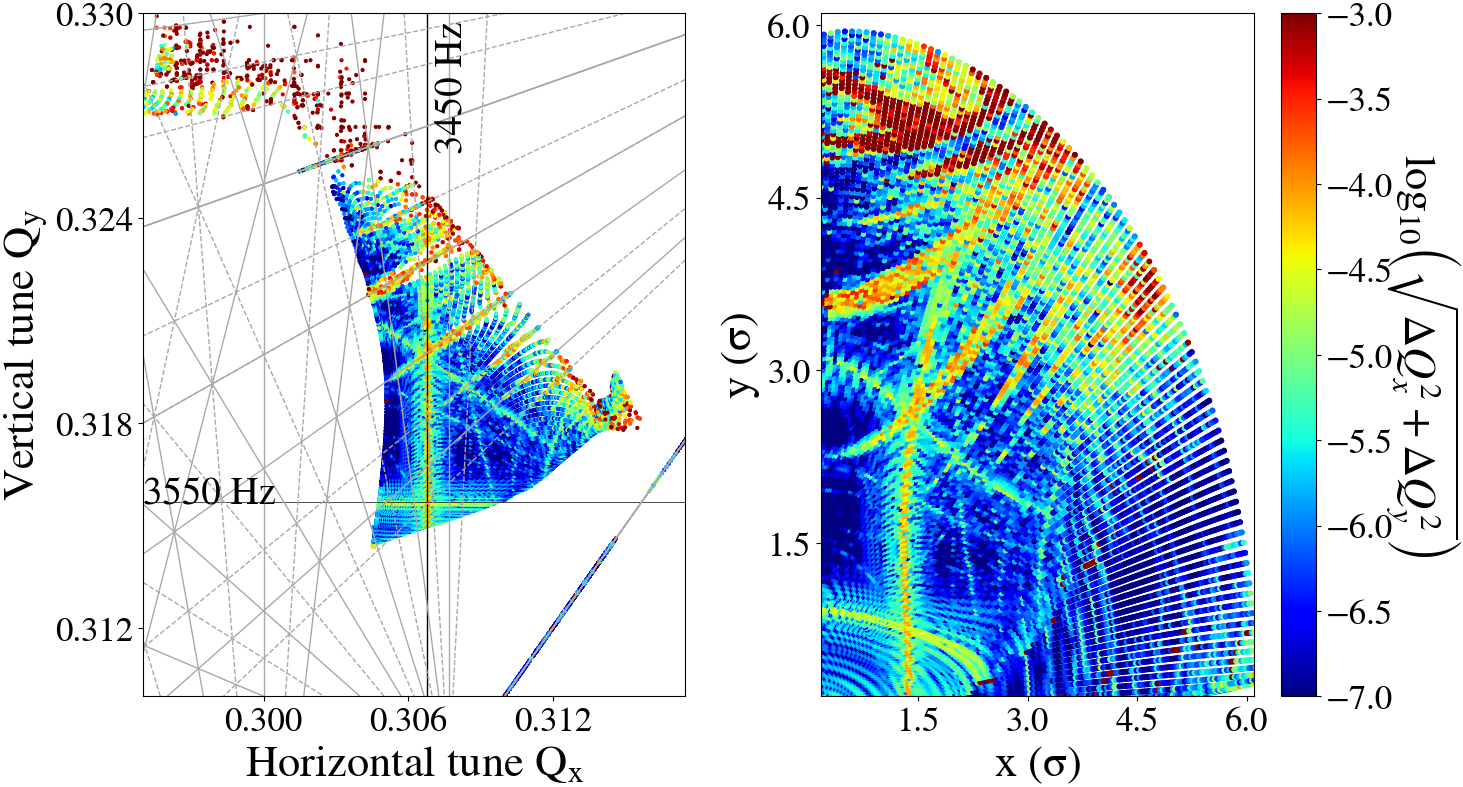} \label{subfig:FMA_low}} \\
\subfloat{\subfigimg[width = 0.49\textwidth]{\textbf{c)}}{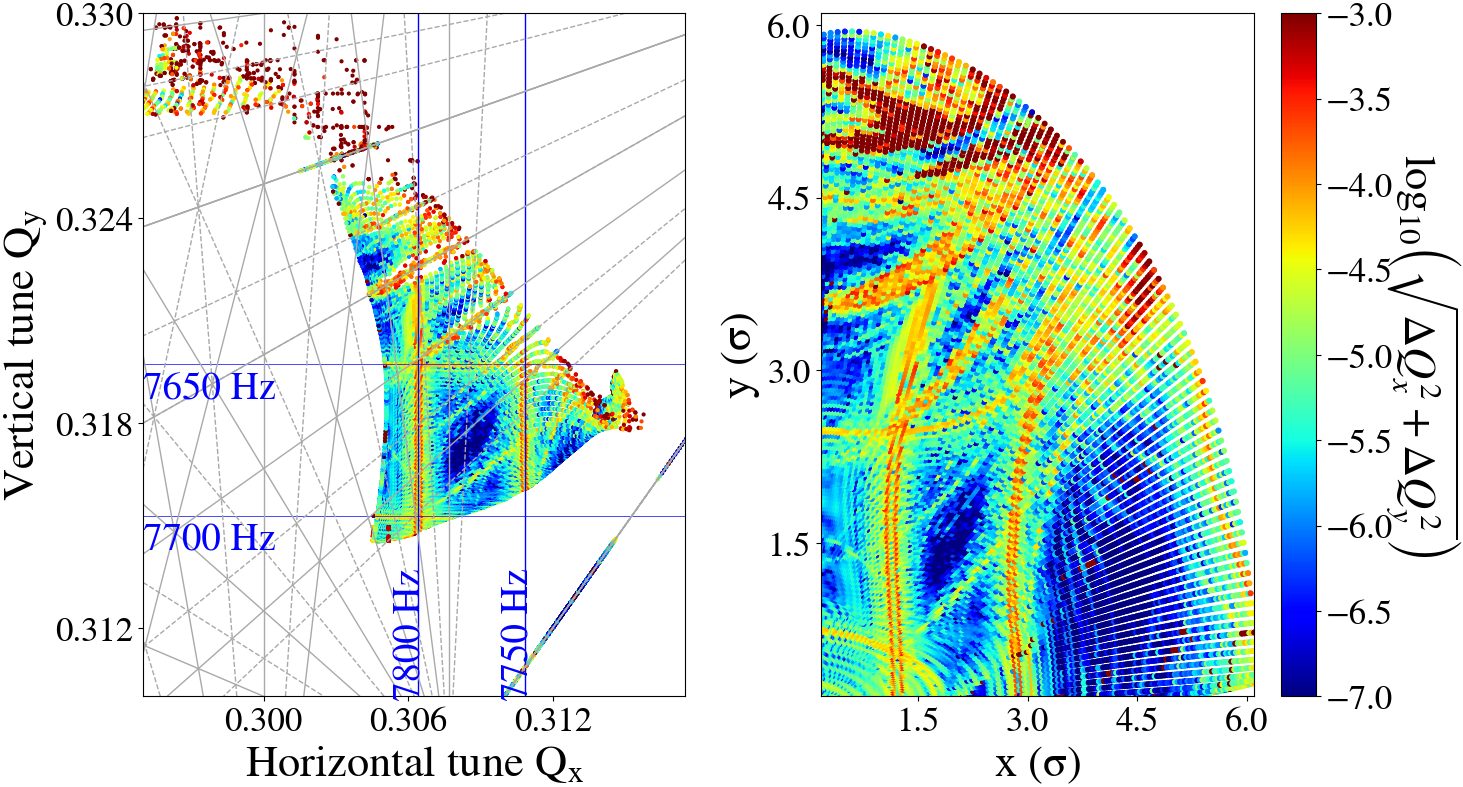} \label{subfig:FMA_high}} 
\subfloat{\subfigimg[width = 0.49\textwidth]{\textbf{d)}}{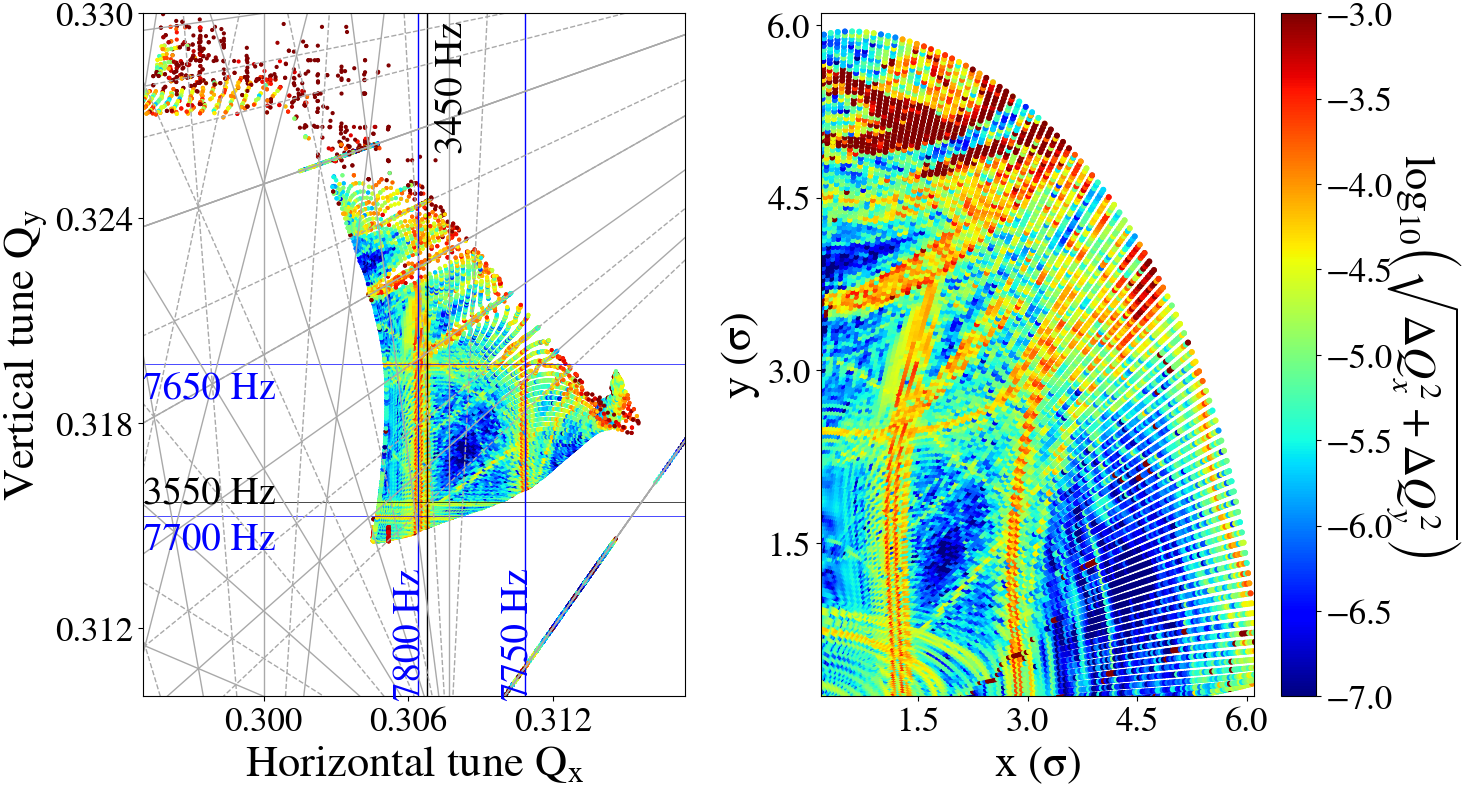} \label{subfig:FMA_all}} \\
\caption{\label{fig:fmas_LHC} The frequency maps (left) and the initial configuration space (right) color-coded with the logarithm of the tune diffusion rate (Eq.~\eqref{eq:diffusion}) for the LHC simulation parameters of Table~\ref{tab:table_parameters_collision} (a) in the absence of power supply ripple , (b) with the 50~Hz harmonics of the low and (c) high-frequency cluster and (d) combining both regimes. The gray lines represent the nominal resonances and the black and blue lines illustrate the resonances (Eq.~\eqref{eq:low} and \eqref{eq:high}) excited due to power supply ripple.}
\end{figure*}

Figure~\ref{fig:fmas_LHC} illustrates the frequency maps (left panel) and the initial configuration space (right panel) for the four studies in the LHC. For the frequency map, the particle tunes of the second interval are plotted and a color code is assigned to logarithm of the tune diffusion rate $D$ to distinguish the stable particles (blue) from the ones with an important variation of their tune (red) due to the resonances. The gray lines denote the nominal resonances, i.e., the resonances that intrinsically arise from the non-linear fields such as non-linear magnets and beam-beam effects without power supply ripple.

In the absence of power supply ripple (Fig.~\ref{subfig:FMA_without}), an important impact is observed due to the third-order resonance ($\text{3} \cdot Q_y \rm =\text{1}$), which is in the vicinity of tune footprint and it affects particles at large amplitudes (red). 

From the FMAs, it is observed that the dipolar power supply ripple results in an increase of the particles' diffusion rate, first, by enhancing the strength of some of the nominal resonances and second, through the excitation of resonances in addition to the nominal ones as described in Eq.~\eqref{eq:tune}. Including the low-frequency cluster (Fig.~\ref{subfig:FMA_low}), the resonances with the largest impact are:
\begin{equation}
\label{eq:low}
    Q_x = Q_p.
\end{equation}
As the power supply ripple is injected in the horizontal plane, these resonances appear as vertical lines in the tune domain in a location equal to the excitation frequency (black). Due to the coupling of the transverse planes, the strongest dipolar excitations are also visible in the vertical plane, appearing as horizontal lines in the frequency maps.

The excitations of the high-frequency cluster appear as aliases (blue). For instance, the excitation at 7.8~kHz in Fig.~\ref{subfig:FMA_high} is located at $f_{\rm rev} \rm -7.8~kHz$. For the high-frequency cluster (Fig.~\ref{subfig:FMA_high}), the most critical resonances are:
\begin{equation}
\label{eq:high}
    Q_x = 1-Q_p.
\end{equation}

For the 50~Hz harmonics in the vicinity of the betatron tune and its alias these additional resonances are located inside the beam's footprint. As clearly shown in the $x-y$ plane (right panel), the existence of such resonances impacts both the core and the tails of the distribution. Reviewing the impact of both clusters (Fig.~\ref{subfig:FMA_all}) indicates that the main contributors to the increase of tune diffusion rate are the spectral components in the high-frequency cluster. Similar results are obtained for the HL-LHC case.


\subsection{Intensity evolution simulations}
\label{sec:losses}
Quantifying the impact of the power supply ripple and the non-linearities on the intensity evolution requires tracking a distribution with realistic profiles in all planes that are similar to the ones observed experimentally. If impacted by resonances, particles at the tails of the distribution close to the limit of DA diffuse and will eventually be lost. Therefore, a detailed representation of the tails of the distribution is needed along with a realistic longitudinal distribution that extends over the whole bucket height. 

The initial conditions form a 4D round distribution, which is sampled from a uniform distribution, extending up to 7 \(\rm \sigma \) both in the initial configuration space, as well as in the initial phase space. In the longitudinal plane, the relative momentum deviation of the particles is a uniform distribution that extends up to the limit of the bucket height (not the value presented in Table~\ref{tab:table_parameters_collision}). 

To reduce the statistical variations, $\rm 25 \times 10^{4}$ particles are tracked in the LHC lattice including synchrotron oscillations. In the post-processing, a weight is assigned to each particle according to its initial amplitude, as computed from the Probability Density Function (PDF) of the final distribution that needs to be simulated. The weight of each particle is:
\begin{equation}
    w = \frac{\prod_{j=1}^{3} \text{PDF}(r_j, \sigma_j) }{\sum_i w_i}, 
\end{equation}
where $j$ iterates over the three planes, $r_j = x_{j}^2 + p_{j}^2$ with $x_j, p_j$ the initial normalized position and momentum coordinates, $\sigma_j$ represents the RMS beam size in the $j$-plane and the denominator denotes the normalization with the sum of the weights of all the particles. 

Experimental observations have shown that the tails of the LHC bunch profiles are overpopulated in the transverse plane and underpopulated in the longitudinal plane compared to a normal distribution \cite{profiles, papadopoulou2018impact}. For an accurate description of the bunch profiles, the PDF of the q-Gaussian distribution is employed \cite{tsallis1, tsallis2, wolfram}:

\begin{equation}
\label{eq:pdf}
\begin{array}{cc}
\text{PDF}(r, \sigma) = \Bigg\{ & 
\begin{array}{cc}
 \frac{e^{-\frac{r^2}{2 \sigma ^2}}}{\sqrt{2 \pi } \sigma } & q=1 \\
 \frac{\sqrt{q-1} \Gamma \left(\frac{1}{q-1}\right) \left(\frac{(q-1) r^2}{2 \sigma ^2}+1\right)^{\frac{1}{1-q}}}{\sqrt{2 \pi } \sigma  \Gamma \left(-\frac{q-3}{2 (q-1)}\right)} & 1<q<3 \\
 \frac{\sqrt{1-q} \Gamma \left(\frac{3}{2}+\frac{1}{1-q}\right) \left(\frac{(q-1) r^2}{2 \sigma ^2}+1\right)^{\frac{1}{1-q}}}{\sqrt{2 \pi } \sigma  \Gamma \left(1+\frac{1}{1-q}\right)} & q<1
\end{array}
 \\
\end{array}
\end{equation}
The parameter $q$ and the ratio of the RMS beam size of the Gaussian ($q$=1) to the q-Gaussian distribution $\sigma_{\rm G} / \sigma_{\rm qG}$ are presented in Table \ref{tab:qgauss} for the longitudinal and the transverse plane (assuming the same profiles for the horizontal and vertical plane).

\begin{table}
\caption{\label{tab:qgauss}%
The parameters of the q-Gaussian PDF shown in Eq.~\eqref{eq:pdf} for the longitudinal and transverse plane \cite{profiles, papadopoulou2018impact}.
}
\begin{ruledtabular}
\begin{tabular}{ccc}
\textrm{Parameters}&
\textrm{Longitudinal plane} & \textrm{Transverse plane}\\
\colrule
 $q$ & 0.88 & 1.15 \\
 $\sigma_{\rm G} / \sigma_{\rm qG}$ & 0.95 & 1.05
\end{tabular}
\end{ruledtabular}
\end{table}
Furthermore, a mechanical aperture is defined in the post-processing at 6.6 \(\rm \sigma\) to simulate the effect of the primary collimators on the transverse plane based on the settings of the 2018 operation. Particles beyond this threshold are considered lost and their weight is set to zero. 

\begin{figure}
\includegraphics[width = \columnwidth]{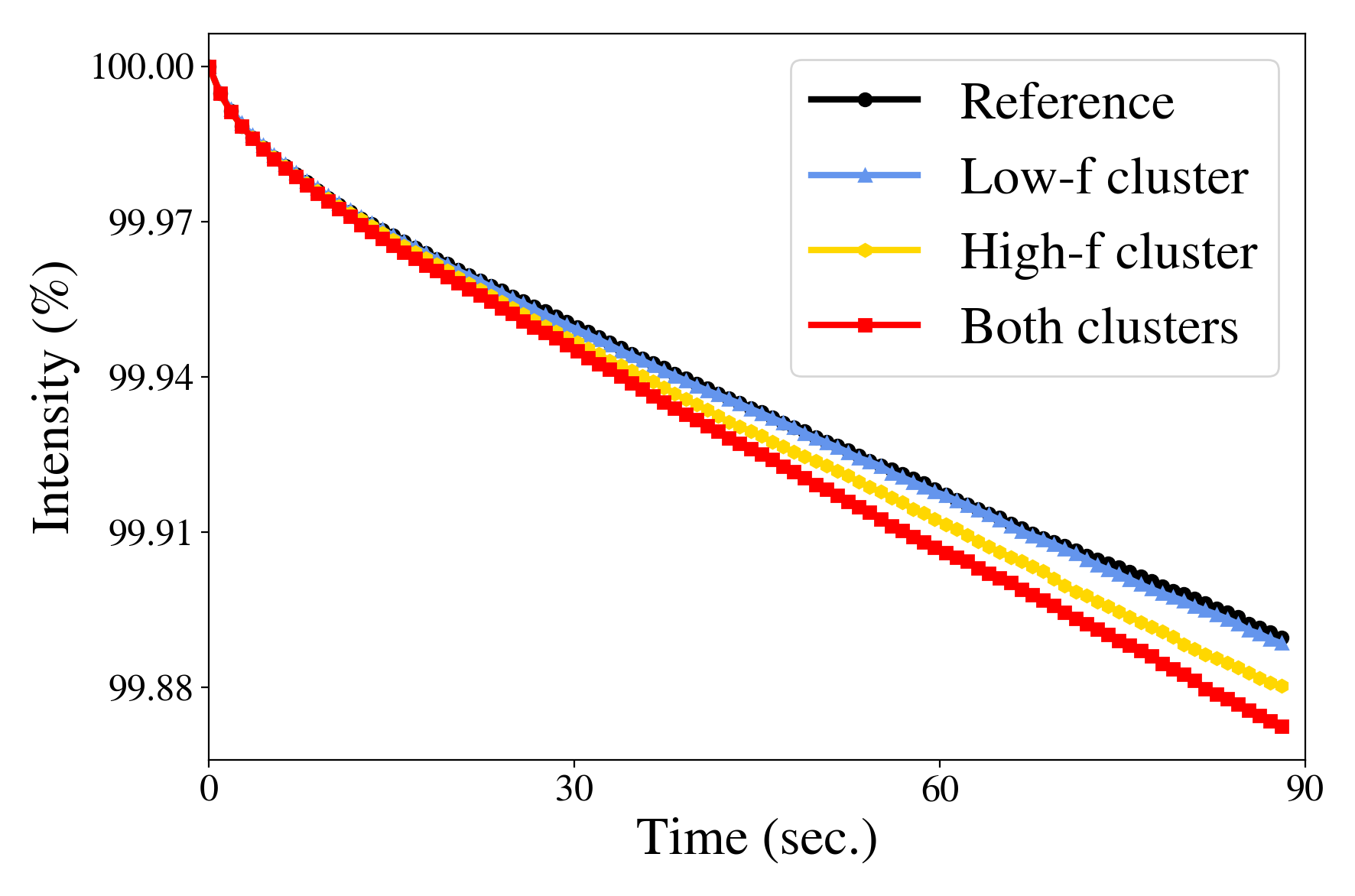}
\caption{\label{fig:losses} Intensity evolution without power supply ripple (black), considering only the low (blue) or high (orange) frequency cluster and including the 50~Hz harmonics in both regimes (red).}
\end{figure}

Figure~\ref{fig:losses} presents the intensity evolution without power supply ripple (black), including the 50~Hz harmonics of the low (blue) or high (orange) frequency cluster and considering both regimes (red). The results indicate that the presence of the 50~Hz harmonics leads to a reduction of the beam intensity, which is already visible with a tracking equivalent to 90 seconds of beam collisions. The results also show that, for the time span under consideration, the high-frequency cluster is the main contributor to the decrease of the beam intensity. A similar effect is observed when considering the same power supply ripple spectrum for the HL-LHC case with the high-frequency cluster acting as the main contributor. 

To quantify the impact of the 50~Hz harmonics on the beam performance, the lifetime $\tau$ is computed from the intensity evolution $I$ as:
\begin{equation}
\label{eq:lifetime}
    I(t) = I_0 \cdot e^{-\frac{t}{\tau}},
\end{equation} 
where $I_0$ is the initial intensity and $t$ the time. An approximation of the instantaneous lifetime is estimated by fitting the exponential decay of the intensity with a sliding window, with each step consisting of a few thousand turns. 

\begin{figure}
\includegraphics[width = \columnwidth]{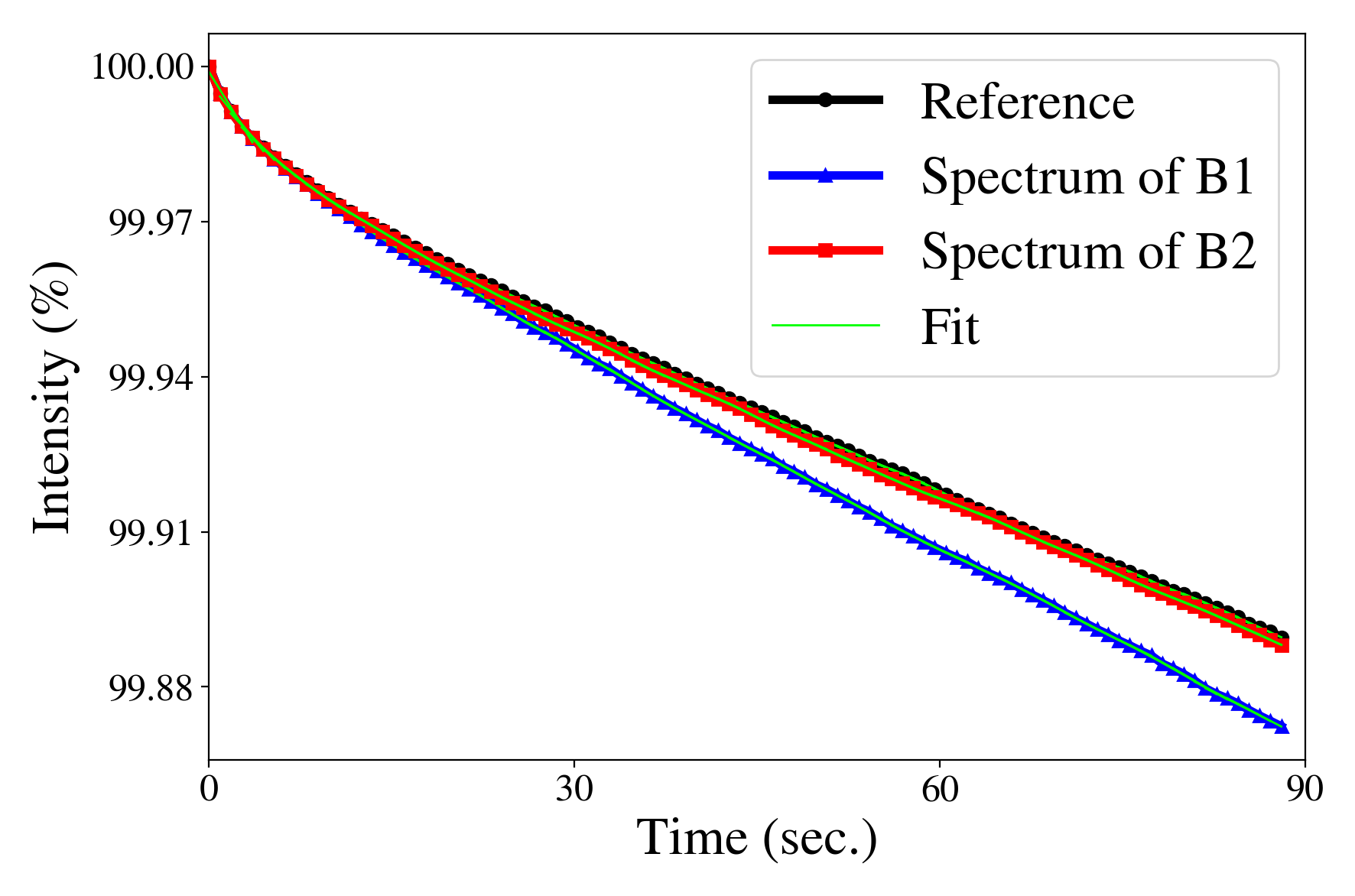}
\caption{\label{fig:losses_b1b2} Intensity evolution in the absence of power supply ripple (black), including the power supply ripple spectrum of Beam 1 (blue) and 2 (red). The green line indicates the fits of the exponential decay of the intensity to compute lifetime (Eq.~\eqref{eq:lifetime}).}
\end{figure}

Figure~\ref{fig:losses_b1b2} illustrates the intensity evolution in the absence of power supply ripple (black), including the power supply ripple spectrum from the experimental observations of Beam 1 (blue) and 2 (red). The green line indicates the fits of Eq.~\eqref{eq:lifetime} to compute the lifetime evolution. 

As shown in Table~\ref{tab:50Hz}, the fact that the power supply ripple spectrum of Beam 2 is lower by approximately a factor of two compared to Beam 1 results in an asymmetry of the intensity evolution between the two beams. Starting from $\tau=28.4$~h in the absence of power supply ripple, the lifetime reduces to $\tau=27.4$~h when including the ripple spectrum of Beam 2 in the simulations and $\tau=22.3$~h when considering the spectrum of Beam 1. During the operation of the accelerator in run 2 (2015-2018), it has been observed that the lifetime of Beam 1 was systematically lower than the lifetime of Beam 2 \cite{b1life1}. It is the first time that simulations reveal that, amongst other mechanisms, the 50~Hz harmonics can contribute to this effect.

\section{Simulation benchmark with controlled excitations}
\label{Sec:adt_excitations}

During the latest LHC run in 2018, controlled dipolar excitations were applied on the beam using the transverse damper kicker. The goal of the experiment was to study the impact of dipolar excitations at various frequencies and amplitudes and to validate our simulation framework in a controlled manner. The experiments were performed at injection energy. 

\begin{figure}
\includegraphics[width = \columnwidth]{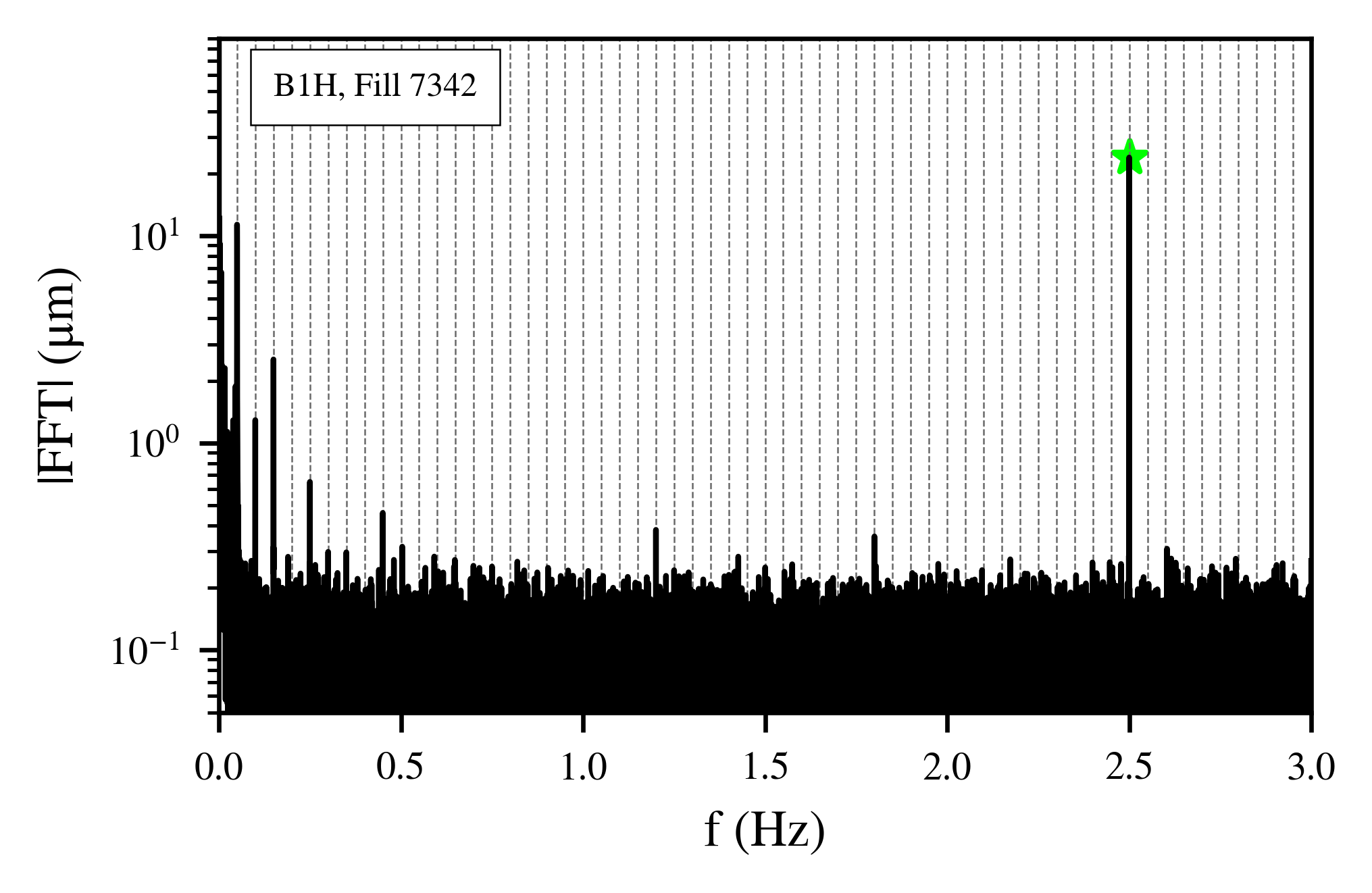}
\caption{\label{fig:excitation_2.5kHz} The horizontal spectrum of a single bunch (black) during a controlled dipolar excitation at 2.5~kHz (green marker) performed using the transverse damper. The vertical lines represent the multiples of 50~Hz.}
\end{figure}

Experimentally, some of the excitations led to a significant reduction of the beam lifetime. To retrieve the initial deflection applied from the transverse damper, the offset and the frequency are extracted from the calibrated ADTObsBox beam spectrum. For instance, Fig.~\ref{fig:excitation_2.5kHz} shows the horizontal spectrum of Beam 1 during a controlled excitation at 2.5~kHz (green star-shaped marker) for a single bunch. The gray vertical lines illustrate the multiples of 50~Hz. Then, using the offset and the frequency, the equivalent kick is computed from Eq.~\ref{appendix:eq_6}. This procedure is repeated for all the excitations applied during the experiments. 

The impact of the excitations on the beam lifetime is compared against the DA scans computed with tracking studies in the presence of dipolar power supply ripple. The important parameters of the tracking simulations at injection energy are summarized in Table~\ref{tab:injection_parameters}.

\begin{table}
\caption{\label{tab:injection_parameters}%
The LHC parameters at injection energy used in the simulations with dipolar power supply ripple.
}
\begin{ruledtabular}
\begin{tabular}{cc}
\textrm{Parameters (unit)}&
\textrm{Values} \\
\colrule
Beam energy (GeV) & 450 \\
Bunch spacing (ns) & 25 \\
RMS bunch length (cm) & 13 \\
RF voltage (MV) &  8 \\
Betatron tunes (\(Q _x, Q_y\)) & (0.275, 0.295) \\
H/V normalized emittance (\(\rm \mu m \ rad\)) & 2.5 \\
H/V chromaticity  & 15 \\
Octupole current (A)  & 20 \\
Bunch population (protons)  & $1.15\times 10^{11}$ \\
Horizontal \(\beta\)-function at Q7 (m)  & 130.9 \\
\end{tabular}
\end{ruledtabular}
\end{table}

In each study, a different combination of the excitation frequency and amplitude is selected. In particular, considering a constant excitation frequency, the value of the deflection is increased and the minimum DA is computed for each case. A ripple amplitude threshold is defined as a function of the excitation frequency. Beyond this limit, a reduction of DA is expected.

\begin{figure}
\includegraphics[width =\columnwidth]{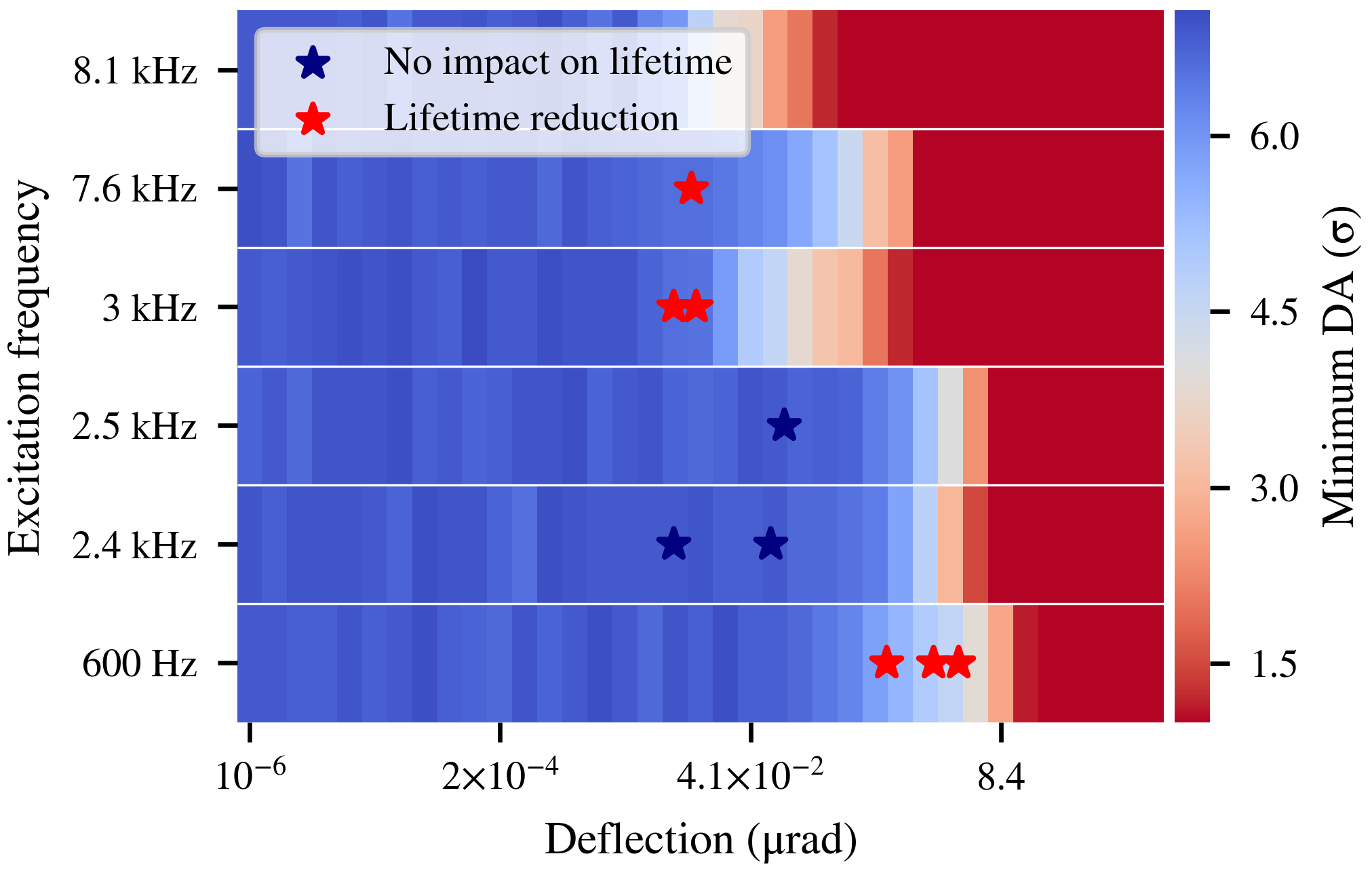} 
\caption{\label{fig:heat_injection} The frequency of the excitation as a function of the deflection. A color code is assigned to the minimum DA computed with tracking simulations including power supply ripple. The star-shaped markers present the equivalent kicks, as computed from the beam spectrum and Eq.~\eqref{appendix:eq_6} during the controlled excitations with the transverse damper. The red markers indicate that the excitation had an impact on the beam lifetime. The blue markers denote the excitations that did not affect the beam lifetime.}
\end{figure}

Figure~\ref{fig:heat_injection} presents the frequency of the excitation as a function of the deflection. A color code is assigned to the minimum DA to distinguish the regime where the power supply ripple has no significant impact (blue) from the one where a significant reduction of DA (red) is observed in the simulations.

In Fig.~\ref{fig:heat_injection}, the star-shaped markers denote the experimental excitation kicks and frequencies. A color code is assigned to the markers that allows distinguishing the excitations that had no impact on lifetime experimentally (blue) from those that lead to a lifetime drop (red). Although an excitation at 8.1~kHz was performed experimentally, the position measurements at this time were not stored and a star-shaped marker is not included. 

The comparison between experimental observations and the power supply ripple threshold defined by DA simulations yields a fairly good agreement between the two for the majority of the excitations, taking into account the simplicity of the machine model and the absence of effects such as linear and non-linear imperfections in the simulations. This comparison provides a validation of our simulation framework including power supply ripple. The method to benchmark simulations and experimental findings presented in this section is not only limited to the studies of the 50~Hz harmonics but can be used to validate the tracking results for different types of power supply ripple and noise effects.

\section{Conclusions}
The purpose of the current paper was to determine whether the 50~Hz harmonics perturbation is a mechanism that can impact the beam performance during the LHC operation. To this end, single-particle tracking simulations were performed in the LHC and HL-LHC lattice including important non-linear fields such as head-on and long-range beam-beam interactions, chromatic sextupoles and Landau octupoles. 

Including a realistic power supply ripple spectrum in the simulations shows that the 50~Hz harmonics increase the tune diffusion of the particles through the excitation of additional resonances, a mechanism that was clearly shown with frequency maps. The high-frequency cluster is identified as the main contributor. 

From simulations that correspond to 90 seconds of beam collisions, the increase of the tune diffusion due to the 50~Hz harmonics results in a lifetime reduction of 21\% with respect to the lifetime in the absence of power supply ripple. Based on these results, it is concluded that the 50~Hz harmonics have an impact on the beam performance during the LHC operation. Mitigation measures should be incorporated in the future operation of the accelerator to suppress the high-frequency cluster from the beam motion.

Due to the asymmetry of the power supply ripple spectrum between Beam 1 and 2 by a factor of approximately two, the simulations illustrate a clear discrepancy in the intensity evolution of the two beams. Including realistic bunch profiles, the estimated lifetime is $\tau=22.3$~h and $\tau=27.4$~h for Beam 1 and 2, respectively. An important lifetime asymmetry between the two beams has been observed since the beginning of run 2 and it is the first time that tracking simulations show that power supply ripple can contribute to this effect.

In the context of this study, a general simulation framework has been developed. This paper illustrated a method to define an acceptable power supply ripple threshold for operation through DA scans. The results of these scans were bench-marked against experimental observations with controlled excitations using the transverse damper. Finally, a method to determine the intensity evolution with weighted distributions and realistic bunch profiles has been demonstrated. The analysis methods presented in this paper can be applied to studies of other types of power supply ripple and noise effects.


\begin{acknowledgments}
The authors gratefully acknowledge H.~Bartosik, M.~C.~Bastos, O.~S.~Brüning, R.~T.~Garcia, M.~Martino and S.~Papadopoulou for valuable suggestions and discussions on this work. We would like to thank D.~Valuch and M.~Soderen for all transverse damper related measurements and experiments.
\end{acknowledgments}

\appendix

\section{Benchmark of analytical formalism for a modulated dipolar field error with tracking simulations}

A comparison between the results of tracking simulations and Eq.~\eqref{appendix:eq_6} is performed as a sanity check. A single particle is tracked in the LHC lattice with SixTrack, in the presence of a dipolar modulation described by Eq.\eqref{eq:mod}. The amplitude of the kick is 1~nrad and the frequency varies across the studies. The offset is computed from the particle's spectrum for each study and is then compared to the analytical formula (Eq.~\eqref{appendix:eq_6}).

\begin{figure}
\includegraphics[width = \columnwidth]{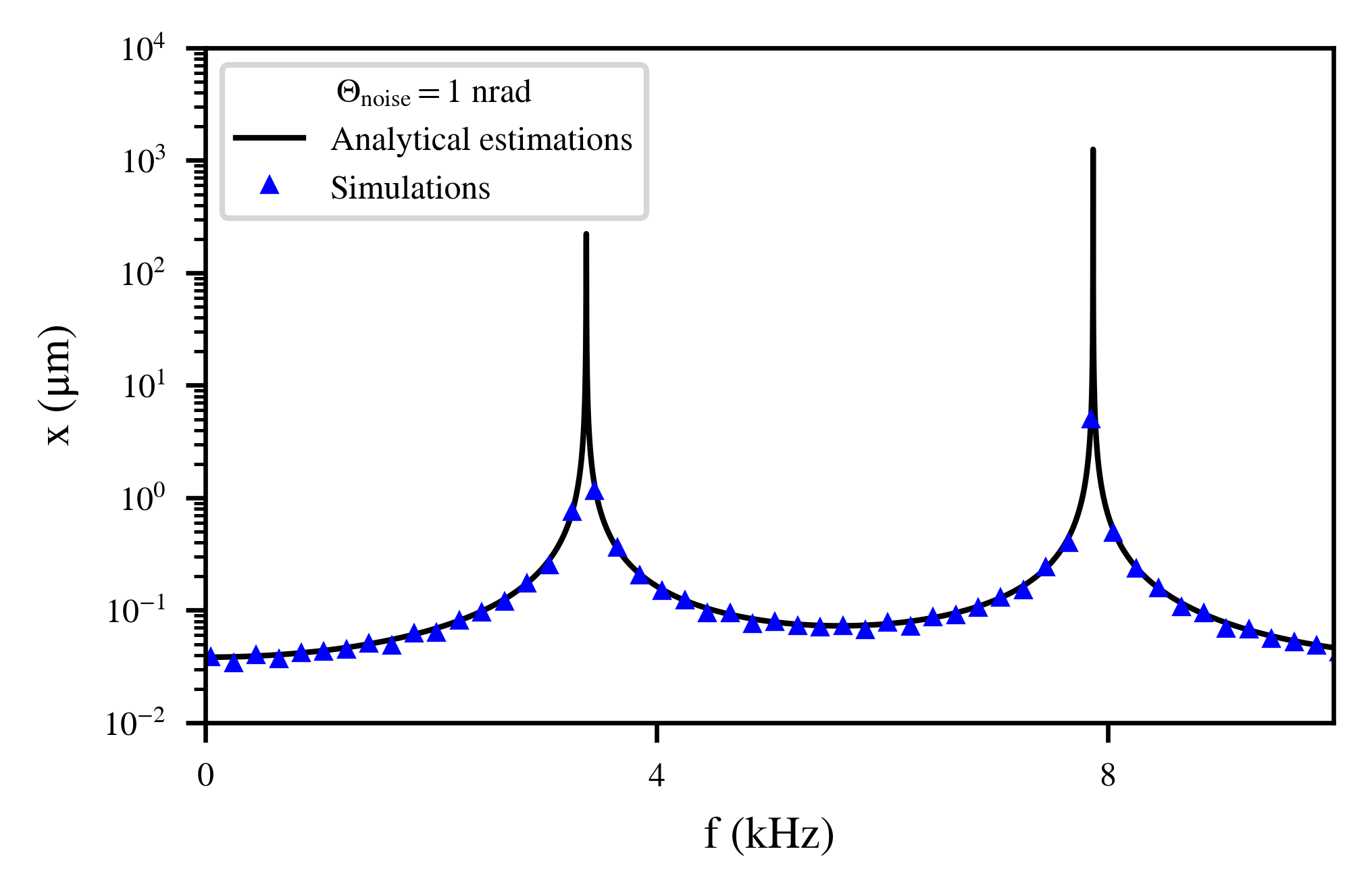}
\caption{\label{fig:appendix:analytical_vs_simulations} The offset as a function of the dipolar excitation frequency with \(\theta_p=\rm 1 \ nrad\) as computed from the closed form of Eq.~\eqref{appendix:eq_6} (black) and from simulations (blue). }
\end{figure}

Figure~\ref{fig:appendix:analytical_vs_simulations} illustrates the offset as a function of the frequency computed analytically (black) and from simulations (blue) and a very good agreement is found between the two. For a constant excitation amplitude, a resonant behavior is expected as the frequency approaches to \(k \cdot f_{\rm rev} \pm f_x\), where \(k\) is an integer. 

As a reference, the maximum offset observed in the beam spectrum is equal to \(\rm 0.84 \times 10^{-3} \sigma \) for a normalized emittance of \(\epsilon_n = \rm 2 \ \mu m \ rad\), a beam energy of 6.5~TeV and a \(\beta-\)function equal to 105 m. Considering a single dipolar excitation at the location of the observation point, the equivalent kick, as computed from Eq.~\eqref{appendix:eq_6}, is \(\theta_p=\rm 0.09~nrad\) for \(Q= \rm 0.31\) and \( |Q-Q_p|=\rm 5\times 10^{-3}\).

\FloatBarrier
\bibliography{bibliography}

\providecommand{\noopsort}[1]{}\providecommand{\singleletter}[1]{#1}%
\begin{thebibliography}{29}%
\makeatletter
\providecommand \@ifxundefined [1]{%
 \@ifx{#1\undefined}
}%
\providecommand \@ifnum [1]{%
 \ifnum #1\expandafter \@firstoftwo
 \else \expandafter \@secondoftwo
 \fi
}%
\providecommand \@ifx [1]{%
 \ifx #1\expandafter \@firstoftwo
 \else \expandafter \@secondoftwo
 \fi
}%
\providecommand \natexlab [1]{#1}%
\providecommand \enquote  [1]{``#1''}%
\providecommand \bibnamefont  [1]{#1}%
\providecommand \bibfnamefont [1]{#1}%
\providecommand \citenamefont [1]{#1}%
\providecommand \href@noop [0]{\@secondoftwo}%
\providecommand \href [0]{\begingroup \@sanitize@url \@href}%
\providecommand \@href[1]{\@@startlink{#1}\@@href}%
\providecommand \@@href[1]{\endgroup#1\@@endlink}%
\providecommand \@sanitize@url [0]{\catcode `\\12\catcode `\$12\catcode
  `\&12\catcode `\#12\catcode `\^12\catcode `\_12\catcode `\%12\relax}%
\providecommand \@@startlink[1]{}%
\providecommand \@@endlink[0]{}%
\providecommand \url  [0]{\begingroup\@sanitize@url \@url }%
\providecommand \@url [1]{\endgroup\@href {#1}{\urlprefix }}%
\providecommand \urlprefix  [0]{URL }%
\providecommand \Eprint [0]{\href }%
\providecommand \doibase [0]{https://doi.org/}%
\providecommand \selectlanguage [0]{\@gobble}%
\providecommand \bibinfo  [0]{\@secondoftwo}%
\providecommand \bibfield  [0]{\@secondoftwo}%
\providecommand \translation [1]{[#1]}%
\providecommand \BibitemOpen [0]{}%
\providecommand \bibitemStop [0]{}%
\providecommand \bibitemNoStop [0]{.\EOS\space}%
\providecommand \EOS [0]{\spacefactor3000\relax}%
\providecommand \BibitemShut  [1]{\csname bibitem#1\endcsname}%
\let\auto@bib@innerbib\@empty
\bibitem [{\citenamefont {{H.~Thiesen and D.~Nisbet}}(2008)}]{SCR}%
  \BibitemOpen
  \bibfield  {author} {\bibinfo {author} {\bibnamefont {{H.~Thiesen and
  D.~Nisbet}}},\ }\bibfield  {title} {\bibinfo {title} {{Review of the initial
  phases of the {LHC} power converter commissioning}},\ }\bibfield  {booktitle}
  {\emph {\bibinfo {booktitle} {{Particle accelerator. Proceedings, 11th
  European Conference, EPAC 2008, Genoa, Italy, June 23-27, 2008}}},\
  }\href@noop {} {\bibfield  {journal} {\bibinfo  {journal} {Conf. Proc.}\
  }\textbf {\bibinfo {volume} {C0806233}},\ \bibinfo {pages} {THPP132}
  (\bibinfo {year} {2008})}\BibitemShut {NoStop}%
\bibitem [{\citenamefont {Brüning}\ \emph {et~al.}(2004)\citenamefont
  {Brüning}, \citenamefont {Collier}, \citenamefont {Lebrun}, \citenamefont
  {Myers}, \citenamefont {Ostojic}, \citenamefont {Poole},\ and\ \citenamefont
  {Proudlock}}]{SCR2}%
  \BibitemOpen
  \bibfield  {author} {\bibinfo {author} {\bibfnamefont {O.~S.}\ \bibnamefont
  {Brüning}}, \bibinfo {author} {\bibfnamefont {P.}~\bibnamefont {Collier}},
  \bibinfo {author} {\bibfnamefont {P.}~\bibnamefont {Lebrun}}, \bibinfo
  {author} {\bibfnamefont {S.}~\bibnamefont {Myers}}, \bibinfo {author}
  {\bibfnamefont {R.}~\bibnamefont {Ostojic}}, \bibinfo {author} {\bibfnamefont
  {J.}~\bibnamefont {Poole}}, and\ \bibinfo {author} {\bibfnamefont
  {P.}~\bibnamefont {Proudlock}},\ }\href
  {https://doi.org/10.5170/CERN-2004-003-V-1} {\emph {\bibinfo {title} {{{LHC}
  Design Report}}}},\ CERN Yellow Reports: Monographs\ (\bibinfo  {publisher}
  {CERN},\ \bibinfo {address} {Geneva},\ \bibinfo {year} {2004})\ Chap.\
  \bibinfo {chapter} {Power converter system}\BibitemShut {NoStop}%
\bibitem [{\citenamefont {Carver}\ \emph {et~al.}(2017)\citenamefont {Carver},
  \citenamefont {Buffat}, \citenamefont {Butterworth}, \citenamefont {Höfle},
  \citenamefont {Iadarola}, \citenamefont {Kotzian}, \citenamefont {Li},
  \citenamefont {Métral}, \citenamefont {Ojeda~Sandonís}, \citenamefont
  {Söderén},\ and\ \citenamefont {Valuch}}]{ADT}%
  \BibitemOpen
  \bibfield  {author} {\bibinfo {author} {\bibfnamefont {L.}~\bibnamefont
  {Carver}}, \bibinfo {author} {\bibfnamefont {X.}~\bibnamefont {Buffat}},
  \bibinfo {author} {\bibfnamefont {A.}~\bibnamefont {Butterworth}}, \bibinfo
  {author} {\bibfnamefont {W.}~\bibnamefont {Höfle}}, \bibinfo {author}
  {\bibfnamefont {G.}~\bibnamefont {Iadarola}}, \bibinfo {author}
  {\bibfnamefont {G.}~\bibnamefont {Kotzian}}, \bibinfo {author} {\bibfnamefont
  {K.}~\bibnamefont {Li}}, \bibinfo {author} {\bibfnamefont {E.}~\bibnamefont
  {Métral}}, \bibinfo {author} {\bibfnamefont {M.}~\bibnamefont
  {Ojeda~Sandonís}}, \bibinfo {author} {\bibfnamefont {M.}~\bibnamefont
  {Söderén}}, and\ \bibinfo {author} {\bibfnamefont {D.}~\bibnamefont
  {Valuch}},\ }\bibfield  {title} {\bibinfo {title} {{Usage of the transverse
  damper observation box for high sampling rate transverse position data in the
  {LHC}}},\ }in\ \href {https://doi.org/10.18429/JACoW-IPAC2017-MOPAB113}
  {\emph {\bibinfo {booktitle} {8th Int. Particle Accelerator Conf.
  (IPAC’17), Copenhagen, Denmark, May 2017}}},\ \bibinfo {series and number}
  {\bibinfo {number} {CERN-ACC-2017-117}}\ (\bibinfo {year} {2017})\ pp.\
  \bibinfo {pages} {389--392}\BibitemShut {NoStop}%
\bibitem [{\citenamefont {Ojeda~Sandonís}\ \emph {et~al.}(2015)\citenamefont
  {Ojeda~Sandonís}, \citenamefont {Baudrenghien}, \citenamefont {Butterworth},
  \citenamefont {Galindo}, \citenamefont {Höfle}, \citenamefont {Levens},
  \citenamefont {Molendijk}, \citenamefont {Vaga},\ and\ \citenamefont
  {Valuch}}]{ADT3}%
  \BibitemOpen
  \bibfield  {author} {\bibinfo {author} {\bibfnamefont {M.}~\bibnamefont
  {Ojeda~Sandonís}}, \bibinfo {author} {\bibfnamefont {P.}~\bibnamefont
  {Baudrenghien}}, \bibinfo {author} {\bibfnamefont {A.}~\bibnamefont
  {Butterworth}}, \bibinfo {author} {\bibfnamefont {J.}~\bibnamefont
  {Galindo}}, \bibinfo {author} {\bibfnamefont {W.}~\bibnamefont {Höfle}},
  \bibinfo {author} {\bibfnamefont {T.}~\bibnamefont {Levens}}, \bibinfo
  {author} {\bibfnamefont {J.}~\bibnamefont {Molendijk}}, \bibinfo {author}
  {\bibfnamefont {F.}~\bibnamefont {Vaga}}, and\ \bibinfo {author}
  {\bibfnamefont {D.}~\bibnamefont {Valuch}},\ }\bibfield  {title} {\bibinfo
  {title} {{Processing high-bandwidth bunch-by-bunch observation data from the
  {RF} and transverse damper systems of the {LHC}}},\ }in\ \href
  {https://doi.org/10.18429/JACoW-ICALEPCS2015-WEPGF062} {\emph {\bibinfo
  {booktitle} {{Proceedings, 15th International Conference on Accelerator and
  Large Experimental Physics Control Systems (ICALEPCS 2015): Melbourne,
  Australia, October 17-23, 2015}}}}\ (\bibinfo {year} {2015})\ p.\ \bibinfo
  {pages} {WEPGF062}\BibitemShut {NoStop}%
\bibitem [{\citenamefont {Söderén}\ \emph {et~al.}(2017)\citenamefont
  {Söderén}, \citenamefont {Kotzian}, \citenamefont {Ojeda~Sandonís},\ and\
  \citenamefont {Valuch}}]{ADT4}%
  \BibitemOpen
  \bibfield  {author} {\bibinfo {author} {\bibfnamefont {M.}~\bibnamefont
  {Söderén}}, \bibinfo {author} {\bibfnamefont {G.}~\bibnamefont {Kotzian}},
  \bibinfo {author} {\bibfnamefont {M.}~\bibnamefont {Ojeda~Sandonís}}, and\
  \bibinfo {author} {\bibfnamefont {D.}~\bibnamefont {Valuch}},\ }\bibfield
  {title} {\bibinfo {title} {{Online bunch by bunch transverse instability
  detection in {LHC}}},\ }in\ \href
  {https://doi.org/10.18429/JACoW-IPAC2017-MOPAB117} {\emph {\bibinfo
  {booktitle} {{Proceedings, 8th International Particle Accelerator Conference
  (IPAC 2017): Copenhagen, Denmark, May 14-19, 2017}}}}\ (\bibinfo {year}
  {2017})\ p.\ \bibinfo {pages} {MOPAB117}\BibitemShut {NoStop}%
\bibitem [{\citenamefont {Carlier}\ \emph {et~al.}(2019)\citenamefont
  {Carlier}, \citenamefont {Tom\'as}, \citenamefont {Maclean},\ and\
  \citenamefont {Persson}}]{tomas1}%
  \BibitemOpen
  \bibfield  {author} {\bibinfo {author} {\bibfnamefont {F.~S.}\ \bibnamefont
  {Carlier}}, \bibinfo {author} {\bibfnamefont {R.}~\bibnamefont {Tom\'as}},
  \bibinfo {author} {\bibfnamefont {E.~H.}\ \bibnamefont {Maclean}}, and\
  \bibinfo {author} {\bibfnamefont {T.~H.~B.}\ \bibnamefont {Persson}},\
  }\bibfield  {title} {\bibinfo {title} {First experimental demonstration of
  forced dynamic aperture measurements with lhc ac dipoles},\ }\href
  {https://doi.org/10.1103/PhysRevAccelBeams.22.031002} {\bibfield  {journal}
  {\bibinfo  {journal} {Phys. Rev. Accel. Beams}\ }\textbf {\bibinfo {volume}
  {22}},\ \bibinfo {pages} {031002} (\bibinfo {year} {2019})}\BibitemShut
  {NoStop}%
\bibitem [{\citenamefont {Tom\'as}(2002)}]{tomas2}%
  \BibitemOpen
  \bibfield  {author} {\bibinfo {author} {\bibfnamefont {R.}~\bibnamefont
  {Tom\'as}},\ }\bibfield  {title} {\bibinfo {title} {Normal form of particle
  motion under the influence of an ac dipole},\ }\href
  {https://doi.org/10.1103/PhysRevSTAB.5.054001} {\bibfield  {journal}
  {\bibinfo  {journal} {Phys. Rev. ST Accel. Beams}\ }\textbf {\bibinfo
  {volume} {5}},\ \bibinfo {pages} {054001} (\bibinfo {year}
  {2002})}\BibitemShut {NoStop}%
\bibitem [{\citenamefont {Apollinari}\ \emph {et~al.}(2017)\citenamefont
  {Apollinari}, \citenamefont {Béjar~Alonso}, \citenamefont {Brüning},
  \citenamefont {Fessia}, \citenamefont {Lamont}, \citenamefont {Rossi},\ and\
  \citenamefont {Tavian}}]{Apollinari:2284929}%
  \BibitemOpen
  \bibfield  {author} {\bibinfo {author} {\bibfnamefont {G.}~\bibnamefont
  {Apollinari}}, \bibinfo {author} {\bibfnamefont {I.}~\bibnamefont
  {Béjar~Alonso}}, \bibinfo {author} {\bibfnamefont {O.}~\bibnamefont
  {Brüning}}, \bibinfo {author} {\bibfnamefont {P.}~\bibnamefont {Fessia}},
  \bibinfo {author} {\bibfnamefont {M.}~\bibnamefont {Lamont}}, \bibinfo
  {author} {\bibfnamefont {L.}~\bibnamefont {Rossi}}, and\ \bibinfo {author}
  {\bibfnamefont {L.}~\bibnamefont {Tavian}},\ }\bibfield  {title} {\bibinfo
  {title} {{High-Luminosity Large Hadron Collider (HL-LHC)}},\ }\href
  {https://doi.org/10.23731/CYRM-2017-004} {\bibfield  {journal} {\bibinfo
  {journal} {CERN Yellow Rep. Monogr.}\ }\textbf {\bibinfo {volume} {4}},\
  \bibinfo {pages} {1} (\bibinfo {year} {2017})}\BibitemShut {NoStop}%
\bibitem [{\citenamefont {Kostoglou}\ \emph {et~al.}(2020)\citenamefont
  {Kostoglou}, \citenamefont {Bartosik}, \citenamefont {Papaphilippou},
  \citenamefont {Sterbini},\ and\ \citenamefont
  {Triantafyllou}}]{kostoglou2020tune}%
  \BibitemOpen
  \bibfield  {author} {\bibinfo {author} {\bibfnamefont {S.}~\bibnamefont
  {Kostoglou}}, \bibinfo {author} {\bibfnamefont {H.}~\bibnamefont {Bartosik}},
  \bibinfo {author} {\bibfnamefont {Y.}~\bibnamefont {Papaphilippou}}, \bibinfo
  {author} {\bibfnamefont {G.}~\bibnamefont {Sterbini}}, and\ \bibinfo {author}
  {\bibfnamefont {N.}~\bibnamefont {Triantafyllou}},\ }\href@noop {} {\bibinfo
  {title} {Tune modulation effects in the high luminosity large hadron
  collider}} (\bibinfo {year} {2020}),\ \Eprint
  {https://arxiv.org/abs/2003.00960} {arXiv:2003.00960 [physics.acc-ph]}
  \BibitemShut {NoStop}%
\bibitem [{six(2019)}]{sixtrack}%
  \BibitemOpen
  \href@noop {} {\bibinfo {title} {Six{T}rack}},\ \bibinfo {howpublished}
  {\url{http://sixtrack.web.cern.ch/SixTrack/}} (\bibinfo {year} {2019}),\
  \bibinfo {note} {accessed: 2019-11-26}\BibitemShut {NoStop}%
\bibitem [{\citenamefont {Maria}\ \emph {et~al.}(2019)\citenamefont {Maria},
  \citenamefont {Andersson}, \citenamefont {Olsen}, \citenamefont {Field},
  \citenamefont {Giovannozzi}, \citenamefont {Hermes}, \citenamefont
  {H{\o}imyr}, \citenamefont {Kostoglou}, \citenamefont {Iadarola},
  \citenamefont {Mcintosh}, \citenamefont {Mereghetti}, \citenamefont {Molson},
  \citenamefont {Pellegrini}, \citenamefont {Persson}, \citenamefont
  {Schwinzerl}, \citenamefont {Maclean}, \citenamefont {Sjobak}, \citenamefont
  {Zacharov},\ and\ \citenamefont {Singh}}]{sixtrack2}%
  \BibitemOpen
  \bibfield  {author} {\bibinfo {author} {\bibfnamefont {R.~D.}\ \bibnamefont
  {Maria}}, \bibinfo {author} {\bibfnamefont {J.}~\bibnamefont {Andersson}},
  \bibinfo {author} {\bibfnamefont {V.~K.~B.}\ \bibnamefont {Olsen}}, \bibinfo
  {author} {\bibfnamefont {L.}~\bibnamefont {Field}}, \bibinfo {author}
  {\bibfnamefont {M.}~\bibnamefont {Giovannozzi}}, \bibinfo {author}
  {\bibfnamefont {P.~D.}\ \bibnamefont {Hermes}}, \bibinfo {author}
  {\bibfnamefont {N.}~\bibnamefont {H{\o}imyr}}, \bibinfo {author}
  {\bibfnamefont {S.}~\bibnamefont {Kostoglou}}, \bibinfo {author}
  {\bibfnamefont {G.}~\bibnamefont {Iadarola}}, \bibinfo {author}
  {\bibfnamefont {E.}~\bibnamefont {Mcintosh}}, \bibinfo {author}
  {\bibfnamefont {A.}~\bibnamefont {Mereghetti}}, \bibinfo {author}
  {\bibfnamefont {J.~W.}\ \bibnamefont {Molson}}, \bibinfo {author}
  {\bibfnamefont {D.}~\bibnamefont {Pellegrini}}, \bibinfo {author}
  {\bibfnamefont {T.}~\bibnamefont {Persson}}, \bibinfo {author} {\bibfnamefont
  {M.}~\bibnamefont {Schwinzerl}}, \bibinfo {author} {\bibfnamefont {E.~H.}\
  \bibnamefont {Maclean}}, \bibinfo {author} {\bibfnamefont {K.}~\bibnamefont
  {Sjobak}}, \bibinfo {author} {\bibfnamefont {I.}~\bibnamefont {Zacharov}},
  and\ \bibinfo {author} {\bibfnamefont {S.}~\bibnamefont {Singh}},\ }\bibfield
   {title} {\bibinfo {title} {{{SixTrack} project: Status, runtime environment,
  and new developments}},\ }in\ \href
  {https://doi.org/10.18429/JACoW-ICAP2018-TUPAF02} {\emph {\bibinfo
  {booktitle} {{Proceedings, 13th International Computational Accelerator
  Physics Conference, ICAP2018: Key West, FL, USA, 20-24 October 2018}}}}\
  (\bibinfo {year} {2019})\ p.\ \bibinfo {pages} {TUPAF02}\BibitemShut
  {NoStop}%
\bibitem [{mad(2019)}]{madx}%
  \BibitemOpen
  \href@noop {} {\bibinfo {title} {{MAD-X}}},\ \bibinfo {howpublished}
  {\url{http://mad.web.cern.ch/mad/}} (\bibinfo {year} {2019}),\ \bibinfo
  {note} {accessed: 2019-12-13}\BibitemShut {NoStop}%
\bibitem [{\citenamefont {Metral}\ \emph {et~al.}(2018)\citenamefont {Metral},
  \citenamefont {Antipov}, \citenamefont {Antoniou}, \citenamefont {Appleby},
  \citenamefont {Arduini}, \citenamefont {Barranco~Garcia}, \citenamefont
  {Baudrenghien}, \citenamefont {Biancacci}, \citenamefont {Bracco},
  \citenamefont {Bruce}, \citenamefont {Buffat}, \citenamefont {Calaga},
  \citenamefont {Carver}, \citenamefont {Chapochnikova}, \citenamefont
  {Crouch}, \citenamefont {De~Maria}, \citenamefont {Fartoukh}, \citenamefont
  {Gamba}, \citenamefont {Giovannozzi}, \citenamefont {Goncalves~Jorge},
  \citenamefont {Hofle}, \citenamefont {Iadarola}, \citenamefont {Karastathis},
  \citenamefont {Lasheen}, \citenamefont {Mastoridis}, \citenamefont
  {Medina~Medrano}, \citenamefont {Mereghetti}, \citenamefont {Mirarchi},
  \citenamefont {Muratori}, \citenamefont {Papadopoulou}, \citenamefont
  {Papaphilippou}, \citenamefont {Pellegrini}, \citenamefont {Pieloni},
  \citenamefont {Redaelli}, \citenamefont {Rumolo}, \citenamefont {Salvant},
  \citenamefont {Solfaroli~Camillocci}, \citenamefont {Tambasco}, \citenamefont
  {Tomas~Garcia},\ and\ \citenamefont {Valuch}}]{HL_LHC_scenarios}%
  \BibitemOpen
  \bibfield  {author} {\bibinfo {author} {\bibfnamefont {E.}~\bibnamefont
  {Metral}}, \bibinfo {author} {\bibfnamefont {S.}~\bibnamefont {Antipov}},
  \bibinfo {author} {\bibfnamefont {F.}~\bibnamefont {Antoniou}}, \bibinfo
  {author} {\bibfnamefont {R.~B.}\ \bibnamefont {Appleby}}, \bibinfo {author}
  {\bibfnamefont {G.}~\bibnamefont {Arduini}}, \bibinfo {author} {\bibfnamefont
  {J.}~\bibnamefont {Barranco~Garcia}}, \bibinfo {author} {\bibfnamefont
  {P.}~\bibnamefont {Baudrenghien}}, \bibinfo {author} {\bibfnamefont
  {N.}~\bibnamefont {Biancacci}}, \bibinfo {author} {\bibfnamefont
  {C.}~\bibnamefont {Bracco}}, \bibinfo {author} {\bibfnamefont
  {R.}~\bibnamefont {Bruce}}, \bibinfo {author} {\bibfnamefont
  {X.}~\bibnamefont {Buffat}}, \bibinfo {author} {\bibfnamefont
  {R.}~\bibnamefont {Calaga}}, \bibinfo {author} {\bibfnamefont {L.~R.}\
  \bibnamefont {Carver}}, \bibinfo {author} {\bibfnamefont {E.}~\bibnamefont
  {Chapochnikova}}, \bibinfo {author} {\bibfnamefont {M.~P.}\ \bibnamefont
  {Crouch}}, \bibinfo {author} {\bibfnamefont {R.}~\bibnamefont {De~Maria}},
  \bibinfo {author} {\bibfnamefont {S.}~\bibnamefont {Fartoukh}}, \bibinfo
  {author} {\bibfnamefont {D.}~\bibnamefont {Gamba}}, \bibinfo {author}
  {\bibfnamefont {M.}~\bibnamefont {Giovannozzi}}, \bibinfo {author}
  {\bibfnamefont {P.}~\bibnamefont {Goncalves~Jorge}}, \bibinfo {author}
  {\bibfnamefont {W.}~\bibnamefont {Hofle}}, \bibinfo {author} {\bibfnamefont
  {G.}~\bibnamefont {Iadarola}}, \bibinfo {author} {\bibfnamefont
  {N.}~\bibnamefont {Karastathis}}, \bibinfo {author} {\bibfnamefont
  {A.}~\bibnamefont {Lasheen}}, \bibinfo {author} {\bibfnamefont
  {T.}~\bibnamefont {Mastoridis}}, \bibinfo {author} {\bibfnamefont {L.~E.}\
  \bibnamefont {Medina~Medrano}}, \bibinfo {author} {\bibfnamefont
  {A.}~\bibnamefont {Mereghetti}}, \bibinfo {author} {\bibfnamefont
  {D.}~\bibnamefont {Mirarchi}}, \bibinfo {author} {\bibfnamefont
  {B.}~\bibnamefont {Muratori}}, \bibinfo {author} {\bibfnamefont {P.~S.}\
  \bibnamefont {Papadopoulou}}, \bibinfo {author} {\bibfnamefont
  {Y.}~\bibnamefont {Papaphilippou}}, \bibinfo {author} {\bibfnamefont
  {D.}~\bibnamefont {Pellegrini}}, \bibinfo {author} {\bibfnamefont
  {T.}~\bibnamefont {Pieloni}}, \bibinfo {author} {\bibfnamefont
  {S.}~\bibnamefont {Redaelli}}, \bibinfo {author} {\bibfnamefont
  {G.}~\bibnamefont {Rumolo}}, \bibinfo {author} {\bibfnamefont
  {B.}~\bibnamefont {Salvant}}, \bibinfo {author} {\bibfnamefont
  {M.}~\bibnamefont {Solfaroli~Camillocci}}, \bibinfo {author} {\bibfnamefont
  {C.}~\bibnamefont {Tambasco}}, \bibinfo {author} {\bibfnamefont
  {R.}~\bibnamefont {Tomas~Garcia}}, and\ \bibinfo {author} {\bibfnamefont
  {D.}~\bibnamefont {Valuch}},\ }\href {http://cds.cern.ch/record/2301292}
  {\emph {\bibinfo {title} {{Update of the {HL-LHC} operational scenarios for
  proton operation}}}},\ \bibinfo {type} {Tech. Rep.}\ (\bibinfo  {institution}
  {CERN},\ \bibinfo {year} {2018})\BibitemShut {NoStop}%
\bibitem [{\citenamefont {Burla}\ \emph {et~al.}(1994)\citenamefont {Burla},
  \citenamefont {Cornuet}, \citenamefont {Fischer}, \citenamefont {Leclère},\
  and\ \citenamefont {Schmidt}}]{Burla:263879}%
  \BibitemOpen
  \bibfield  {author} {\bibinfo {author} {\bibfnamefont {P.}~\bibnamefont
  {Burla}}, \bibinfo {author} {\bibfnamefont {D.}~\bibnamefont {Cornuet}},
  \bibinfo {author} {\bibfnamefont {K.}~\bibnamefont {Fischer}}, \bibinfo
  {author} {\bibfnamefont {P.}~\bibnamefont {Leclère}}, and\ \bibinfo {author}
  {\bibfnamefont {F.}~\bibnamefont {Schmidt}},\ }\href
  {http://cds.cern.ch/record/263879} {\emph {\bibinfo {title} {{Power supply
  ripple study at the {SPS}}}}},\ \bibinfo {type} {Tech. Rep.}\ \bibinfo
  {number} {CERN-SL-94-11-AP}\ (\bibinfo  {institution} {CERN},\ \bibinfo
  {address} {Geneva},\ \bibinfo {year} {1994})\BibitemShut {NoStop}%
\bibitem [{\citenamefont {{{O.~S.~Br{\"u}ning and
  F.~Willeke}}}(1994)}]{intro1994_1}%
  \BibitemOpen
  \bibfield  {author} {\bibinfo {author} {\bibnamefont {{{O.~S.~Br{\"u}ning and
  F.~Willeke}}}},\ }\bibfield  {title} {\bibinfo {title} {{Diffusion-like
  processes in proton storage rings due to the combined effect of non-linear
  fields and modulational effects with more than one frequency}},\ }in\ \href
  {https://cds.cern.ch/record/271555} {\emph {\bibinfo {booktitle} {European
  Particle Accelerator Conference (EPAC 1994)}}}\ (\bibinfo {year}
  {1994})\BibitemShut {NoStop}%
\bibitem [{\citenamefont {Brüning}(1992)}]{SPS2_tune}%
  \BibitemOpen
  \bibfield  {author} {\bibinfo {author} {\bibfnamefont {O.~S.}\ \bibnamefont
  {Brüning}},\ }\bibfield  {title} {\bibinfo {title} {Diffusion in a {FODO}
  cell due to modulation effects in the presence of nonlinear fields},\ }\href
  {https://cds.cern.ch/record/1108270} {\bibfield  {journal} {\bibinfo
  {journal} {Part. Accel.}\ }\textbf {\bibinfo {volume} {41}},\ \bibinfo
  {pages} {133} (\bibinfo {year} {1992})}\BibitemShut {NoStop}%
\bibitem [{\citenamefont {Papaphilippou}(2014)}]{fma1}%
  \BibitemOpen
  \bibfield  {author} {\bibinfo {author} {\bibfnamefont {Y.}~\bibnamefont
  {Papaphilippou}},\ }\bibfield  {title} {\bibinfo {title} {Detecting chaos in
  particle accelerators through the frequency map analysis method},\
  }\href@noop {} {\bibfield  {journal} {\bibinfo  {journal} {Chaos: An
  Interdisciplinary Journal of Nonlinear Science}\ }\textbf {\bibinfo {volume}
  {24}},\ \bibinfo {pages} {024412} (\bibinfo {year} {2014})}\BibitemShut
  {NoStop}%
\bibitem [{\citenamefont {Laskar}(1999)}]{fma2}%
  \BibitemOpen
  \bibfield  {author} {\bibinfo {author} {\bibfnamefont {J.}~\bibnamefont
  {Laskar}},\ }\bibfield  {title} {\bibinfo {title} {Introduction to frequency
  map analysis},\ }in\ \href@noop {} {\emph {\bibinfo {booktitle} {Hamiltonian
  systems with three or more degrees of freedom}}}\ (\bibinfo  {publisher}
  {Springer},\ \bibinfo {year} {1999})\ pp.\ \bibinfo {pages}
  {134--150}\BibitemShut {NoStop}%
\bibitem [{\citenamefont {Laskar}(2003)}]{fma3}%
  \BibitemOpen
  \bibfield  {author} {\bibinfo {author} {\bibfnamefont {J.}~\bibnamefont
  {Laskar}},\ }\bibfield  {title} {\bibinfo {title} {Frequency map analysis and
  particle accelerators},\ }in\ \href@noop {} {\emph {\bibinfo {booktitle}
  {Proceedings of the 2003 Particle Accelerator Conference}}},\ Vol.~\bibinfo
  {volume} {1}\ (\bibinfo {organization} {IEEE},\ \bibinfo {year} {2003})\ pp.\
  \bibinfo {pages} {378--382}\BibitemShut {NoStop}%
\bibitem [{\citenamefont {Laskar}(2000)}]{fma4}%
  \BibitemOpen
  \bibfield  {author} {\bibinfo {author} {\bibfnamefont {J.}~\bibnamefont
  {Laskar}},\ }\bibfield  {title} {\bibinfo {title} {Application of frequency
  map analysis},\ }in\ \href@noop {} {\emph {\bibinfo {booktitle} {The Chaotic
  Universe: Proceedings of the Second ICRA Network Workshop, Rome, Pescara,
  Italy, 1-5 February 1999}}},\ Vol.~\bibinfo {volume} {10}\ (\bibinfo
  {organization} {World Scientific},\ \bibinfo {year} {2000})\ p.\ \bibinfo
  {pages} {115}\BibitemShut {NoStop}%
\bibitem [{\citenamefont {Laskar}\ \emph {et~al.}(1992)\citenamefont {Laskar},
  \citenamefont {Froeschl{\'e}},\ and\ \citenamefont {Celletti}}]{NAFF1}%
  \BibitemOpen
  \bibfield  {author} {\bibinfo {author} {\bibfnamefont {J.}~\bibnamefont
  {Laskar}}, \bibinfo {author} {\bibfnamefont {C.}~\bibnamefont
  {Froeschl{\'e}}}, and\ \bibinfo {author} {\bibfnamefont {A.}~\bibnamefont
  {Celletti}},\ }\bibfield  {title} {\bibinfo {title} {The measure of chaos by
  the numerical analysis of the fundamental frequencies. application to the
  standard mapping},\ }\href@noop {} {\bibfield  {journal} {\bibinfo  {journal}
  {Physica D: Nonlinear Phenomena}\ }\textbf {\bibinfo {volume} {56}},\
  \bibinfo {pages} {253} (\bibinfo {year} {1992})}\BibitemShut {NoStop}%
\bibitem [{\citenamefont {Kostoglou}\ \emph {et~al.}(2017)\citenamefont
  {Kostoglou}, \citenamefont {Karastathis}, \citenamefont {Papaphilippou},
  \citenamefont {Pellegrini},\ and\ \citenamefont {Zisopoulos}}]{NAFF}%
  \BibitemOpen
  \bibfield  {author} {\bibinfo {author} {\bibfnamefont {S.}~\bibnamefont
  {Kostoglou}}, \bibinfo {author} {\bibfnamefont {N.}~\bibnamefont
  {Karastathis}}, \bibinfo {author} {\bibfnamefont {Y.}~\bibnamefont
  {Papaphilippou}}, \bibinfo {author} {\bibfnamefont {D.}~\bibnamefont
  {Pellegrini}}, and\ \bibinfo {author} {\bibfnamefont {P.}~\bibnamefont
  {Zisopoulos}},\ }\bibfield  {title} {\bibinfo {title} {Development of
  computational tools for noise studies in the {LHC}},\ }in\ \href@noop {}
  {\emph {\bibinfo {booktitle} {Conf. Proc.}}},\ \bibinfo {series and number}
  {\bibinfo {number} {IPAC-2017-THPAB044}}\ (\bibinfo {year} {2017})\ p.\
  \bibinfo {pages} {THPAB044}\BibitemShut {NoStop}%
\bibitem [{\citenamefont {Zisopoulos}\ \emph {et~al.}(2019)\citenamefont
  {Zisopoulos}, \citenamefont {Papaphilippou},\ and\ \citenamefont
  {Laskar}}]{PhysRevAccelBeams.22.071002}%
  \BibitemOpen
  \bibfield  {author} {\bibinfo {author} {\bibfnamefont {P.}~\bibnamefont
  {Zisopoulos}}, \bibinfo {author} {\bibfnamefont {Y.}~\bibnamefont
  {Papaphilippou}}, and\ \bibinfo {author} {\bibfnamefont {J.}~\bibnamefont
  {Laskar}},\ }\bibfield  {title} {\bibinfo {title} {Refined betatron tune
  measurements by mixing beam position data},\ }\href
  {https://doi.org/10.1103/PhysRevAccelBeams.22.071002} {\bibfield  {journal}
  {\bibinfo  {journal} {Phys. Rev. Accel. Beams}\ }\textbf {\bibinfo {volume}
  {22}},\ \bibinfo {pages} {071002} (\bibinfo {year} {2019})}\BibitemShut
  {NoStop}%
\bibitem [{\citenamefont {Papadopoulou}\ \emph {et~al.}(2017)\citenamefont
  {Papadopoulou}, \citenamefont {Antoniou}, \citenamefont {Argyropoulos},
  \citenamefont {Fitterer}, \citenamefont {Hostettler},\ and\ \citenamefont
  {Papaphilippou}}]{profiles}%
  \BibitemOpen
  \bibfield  {author} {\bibinfo {author} {\bibfnamefont {S.}~\bibnamefont
  {Papadopoulou}}, \bibinfo {author} {\bibfnamefont {F.}~\bibnamefont
  {Antoniou}}, \bibinfo {author} {\bibfnamefont {T.}~\bibnamefont
  {Argyropoulos}}, \bibinfo {author} {\bibfnamefont {M.}~\bibnamefont
  {Fitterer}}, \bibinfo {author} {\bibfnamefont {M.}~\bibnamefont
  {Hostettler}}, and\ \bibinfo {author} {\bibfnamefont {Y.}~\bibnamefont
  {Papaphilippou}},\ }\bibfield  {title} {\bibinfo {title} {Modelling and
  measurements of bunch profiles at the {LHC}},\ }\href
  {https://doi.org/10.1088/1742-6596/874/1/012008} {\bibfield  {journal}
  {\bibinfo  {journal} {Journal of Physics: Conference Series}\ }\textbf
  {\bibinfo {volume} {874}},\ \bibinfo {pages} {012008} (\bibinfo {year}
  {2017})}\BibitemShut {NoStop}%
\bibitem [{\citenamefont {Papadopoulou}\ \emph {et~al.}(2018)\citenamefont
  {Papadopoulou}, \citenamefont {Antoniou}, \citenamefont {Argyropoulos},
  \citenamefont {Hostettler}, \citenamefont {Papaphilippou},\ and\
  \citenamefont {Trad}}]{papadopoulou2018impact}%
  \BibitemOpen
  \bibfield  {author} {\bibinfo {author} {\bibfnamefont {S.}~\bibnamefont
  {Papadopoulou}}, \bibinfo {author} {\bibfnamefont {F.}~\bibnamefont
  {Antoniou}}, \bibinfo {author} {\bibfnamefont {T.}~\bibnamefont
  {Argyropoulos}}, \bibinfo {author} {\bibfnamefont {M.}~\bibnamefont
  {Hostettler}}, \bibinfo {author} {\bibfnamefont {Y.}~\bibnamefont
  {Papaphilippou}}, and\ \bibinfo {author} {\bibfnamefont {G.}~\bibnamefont
  {Trad}},\ }\href@noop {} {\bibinfo {title} {Impact of non-gaussian beam
  profiles in the performance of hadron colliders}} (\bibinfo {year} {2018}),\
  \Eprint {https://arxiv.org/abs/1806.07317} {arXiv:1806.07317
  [physics.acc-ph]} \BibitemShut {NoStop}%
\bibitem [{\citenamefont {Tsallis}\ \emph {et~al.}(2003)\citenamefont
  {Tsallis}, \citenamefont {Baldovin}, \citenamefont {Cerbino},\ and\
  \citenamefont {Pierobon}}]{tsallis1}%
  \BibitemOpen
  \bibfield  {author} {\bibinfo {author} {\bibfnamefont {C.}~\bibnamefont
  {Tsallis}}, \bibinfo {author} {\bibfnamefont {F.}~\bibnamefont {Baldovin}},
  \bibinfo {author} {\bibfnamefont {R.}~\bibnamefont {Cerbino}}, and\ \bibinfo
  {author} {\bibfnamefont {P.}~\bibnamefont {Pierobon}},\ }\href@noop {}
  {\bibinfo {title} {Introduction to nonextensive statistical mechanics and
  thermodynamics}} (\bibinfo {year} {2003}),\ \Eprint
  {https://arxiv.org/abs/cond-mat/0309093} {arXiv:cond-mat/0309093
  [cond-mat.stat-mech]} \BibitemShut {NoStop}%
\bibitem [{\citenamefont {{E.~M.~F.~Curado and C.~Tsallis}}(1991)}]{tsallis2}%
  \BibitemOpen
  \bibfield  {author} {\bibinfo {author} {\bibnamefont {{E.~M.~F.~Curado and
  C.~Tsallis}}},\ }\bibfield  {title} {\bibinfo {title} {{Generalized
  statistical mechanics: Connection with thermodynamics}},\ }\href@noop {}
  {\bibfield  {journal} {\bibinfo  {journal} {J. Phys. A}\ }\textbf {\bibinfo
  {volume} {24}},\ \bibinfo {pages} {L69} (\bibinfo {year} {1991})},\ \bibinfo
  {note} {[Erratum: J.Phys.A 25, 1019 (1992)]}\BibitemShut {NoStop}%
\bibitem [{\citenamefont {{Wolfram Research{,} Inc.}}()}]{wolfram}%
  \BibitemOpen
  \bibfield  {author} {\bibinfo {author} {\bibnamefont {{Wolfram Research{,}
  Inc.}}},\ }\href
  {https://reference.wolfram.com/language/ref/TsallisQGaussianDistribution.html}
  {\bibinfo {title} {{Tsallis q-Gaussian distribution}}},\ \bibinfo {note}
  {accessed: 2020-06-08}\BibitemShut {NoStop}%
\bibitem [{\citenamefont {Papadopoulou}\ \emph {et~al.}(2019)\citenamefont
  {Papadopoulou}, \citenamefont {Antoniou}, \citenamefont {Efthymiopoulos},
  \citenamefont {Hostettler}, \citenamefont {Iadarola}, \citenamefont
  {Karastathis}, \citenamefont {Kostoglou}, \citenamefont {Papaphilippou},\
  and\ \citenamefont {Trad}}]{b1life1}%
  \BibitemOpen
  \bibfield  {author} {\bibinfo {author} {\bibfnamefont {S.}~\bibnamefont
  {Papadopoulou}}, \bibinfo {author} {\bibfnamefont {F.}~\bibnamefont
  {Antoniou}}, \bibinfo {author} {\bibfnamefont {I.}~\bibnamefont
  {Efthymiopoulos}}, \bibinfo {author} {\bibfnamefont {M.}~\bibnamefont
  {Hostettler}}, \bibinfo {author} {\bibfnamefont {G.}~\bibnamefont
  {Iadarola}}, \bibinfo {author} {\bibfnamefont {N.}~\bibnamefont
  {Karastathis}}, \bibinfo {author} {\bibfnamefont {S.}~\bibnamefont
  {Kostoglou}}, \bibinfo {author} {\bibfnamefont {Y.}~\bibnamefont
  {Papaphilippou}}, and\ \bibinfo {author} {\bibfnamefont {G.}~\bibnamefont
  {Trad}},\ }\bibfield  {title} {\bibinfo {title} {Monitoring and modelling of
  the {LHC} emittance and luminosity evolution in 2018},\ }\href
  {https://doi.org/10.1088/1742-6596/1350/1/012011} {\bibfield  {journal}
  {\bibinfo  {journal} {Journal of Physics: Conference Series}\ }\textbf
  {\bibinfo {volume} {1350}},\ \bibinfo {pages} {012011} (\bibinfo {year}
  {2019})}\BibitemShut {NoStop}%
\end{thebibliography}%


\providecommand{\noopsort}[1]{}\providecommand{\singleletter}[1]{#1}%
%
\end{document}